\documentclass[12pt]{article}
\usepackage{CJK}
\usepackage{CJKnumb}
\usepackage{graphicx}
\usepackage{subfigure}
\usepackage{amsfonts}
\usepackage{amsmath}
\usepackage{amssymb}
\usepackage{fancyhdr}
\usepackage{indentfirst}
\usepackage{setspace}
    \usepackage{titlesec}
    \usepackage{natbib}
\usepackage{hyperref}
\usepackage{cleveref}
\usepackage{color}
\usepackage{colortbl}
\usepackage{hhline}
    \titleformat{\section}{\large\bfseries}{\thesection}{1em}{}
    \titleformat{\subsection}{\large\bfseries}{\thesubsection}{1em}{}
 
\newcommand{\single}{\renewcommand{\baselinestretch}{1.2}\normalsize}
\newcommand{\double}{\renewcommand{\baselinestretch}{1.63}\normalsize}

\newcommand{\rss}{\vspace{-.4cm}}

\newtheorem{theorem}{Theorem}
\newtheorem{lemma}[theorem]{Lemma}
\newtheorem{corollary}[theorem]{Corollary}
\def\R{\mathbb{R}}

\def\cov{{\rm cov}}
\def \E{{\rm E}}
\def\var{{\rm var}}

\newcommand{\bc}{\begin{center}}
\newcommand{\ec}{\end{center}}
\newcommand{\mt}{\mathcal{T}}
\newcommand{\ms}{\mathcal{S}}

\newcommand{\inner}[2]{\langle #1,#2\rangle}
\newcommand{\mT}{\mathcal{T}}
\newcommand{\mS}{\mathcal{S}}

\renewcommand{\baselinestretch}{1.2}
\title{A test of weak separability for multi-way functional data, with application to brain connectivity studies}
\author{Brian Lynch, Kehui Chen \\
University of Pittsburgh}
\begin{document}
\setlength{\belowdisplayskip}{6pt} \setlength{\belowdisplayshortskip}{6pt}
\setlength{\abovedisplayskip}{6pt} \setlength{\abovedisplayshortskip}{6pt}

\pagenumbering{Alph}
\begin{titlepage}
\thispagestyle{empty}
\single
\maketitle

%

\begin{center}
  \textbf{Abstract:}
  \end{center}
This paper concerns the modeling of multi-way functional data where double or multiple indices are involved. We introduce a concept of weak separability. The weakly separable structure supports the use of factorization methods that decompose the signal into its spatial  and temporal  components. The analysis reveals interesting connections to the usual strongly separable covariance structure, and provides insights into tensor methods for multi-way functional data. We propose a formal test for the weak separability hypothesis, where the asymptotic null distribution of the test statistic is a chi-square type mixture. The method is applied to study brain functional connectivity derived from source localized magnetoencephalography signals during motor tasks.

\vspace*{.1in}


\thispagestyle{empty} \vfill
\noindent \vspace{-.2cm}\rule{\textwidth}{0.5pt}\\
{\small Brian Lynch is a PhD Student, Department
of Statistics, University of Pittsburgh, Pittsburgh, PA 15260
(E-mail: bcl28@pitt.edu).}
{\small Kehui Chen is Assistant Professor, Department
of Statistics, University of Pittsburgh, Pittsburgh, PA 15260
(E-mail: khchen@pitt.edu).}
{\small This work is partially supported by NSF1612458.}

\end{titlepage}

\pagenumbering{arabic} 
\double


\rss\rss\section{Introduction}\rss
\label{sec:intro}

Traditional functional data analysis usually concerns data recorded over a continuum, such as growth curves. Dense and regularly-observed functional data can be recorded in a matrix with dimension $n\times T$, where $n$ is the number of subjects and $T$ is the number of grid points observed for each subject. Multi-way functional data refers to an extension where multiple indices are involved and data can be recorded in a tensor with dimension at least three. Examples include brain imaging data where for each subject $i = 1,\ldots, n$, we have observations $X_i(s, t)$, with a spatial index $s \in \R^{d}$ and a time index $t\in \R^1$. Other examples include repeatedly or longitudinally observed functional data, such as data obtained from tracking apps where subjects' 24-hour profiles of activities are recorded every day. This type of data can be represented by $X_i(s,t)$, where $s$ denotes the day and $t$ denotes the time within a day. 

As multi-way functional data become more common with modern techniques, the modeling of this type of data attracts increasing interest.
Assume the individual observations $X_i(s,t)$ are independent and identically distributed realizations of a random process $X \in L^2(\mS\times \mT)$, $s\in \mS\subseteq\R^{d_1}$, $t\in \mT\subseteq\R^{d_2}$, with mean $\mu$ and continuous covariance operator $C$. When we can do so without confusion, we use the same symbol for the covariance operator and its kernel function. A well-established tool in functional data analysis is functional principal component analysis. When applied to the multi-way process $X$, functional principal component analysis is based on the Karhunen--Lo\`eve representation $X(s,t) = \mu(s,t) + \sum_{l=1}^{\infty} Z_l h_l(s,t)$, where $Z_l\ (l= 1,2,\ldots )$ are the (random) uncorrelated coefficients, and $h_l(s,t)\ (l = 1,2,\ldots )$ are the eigenfunctions of the covariance operator $C$.

To alleviate the difficulties associated with modeling the $(2d_1+2d_2)$-dimensional full covariance function $C(s,t,u,v)$ and characterizing the $(d_1+d_2)$-dimensional eigenfunctions, one usually seeks dimension reduction through factorization of the signal into its spatial and temporal components. \cite{chen2015modeling} proposed product functional principal component analysis,
\begin{equation}
	\label{productFPCA}
X(s,t) = \mu(s,t) + \sum_{k=1}^{\infty}\sum_{j=1}^{\infty}\chi_{jk}\psi_j(s)\phi_k(t),
\end{equation}
 where $\psi_j(s)\ (j= 1,2,\ldots )$ and $\phi_k(t)\ (k=1,2,\ldots )$ are the eigenfunctions of the marginal covariance operators in $L^2(\mathcal{S})$ and $L^2(\mathcal{T})$, with corresponding marginal kernels
\begin{equation}
   \label{eq:marginalcov}
   C_{\mathcal{S}}(s,u) = \int_{\mathcal{T}} C(s,t;u,t)dt, \quad C_{\mathcal{T}}(t,v) =\int_{\mathcal{S}} C(s,t;s,v) ds.\end{equation}
Here $C_{\mathcal{S}}(s,u) = \sum_{j=1}^{\infty}\lambda_j\psi_j(s)\psi_j(u)$ and $C_{\mT}(t,v)=\sum_{k=1}^{\infty}\gamma_k\phi_k(t)\phi_k(v)$, where $\lambda_1 \ge \lambda_2 \ge \cdots$ and $\gamma_1 \ge \gamma_2 \ge \cdots$ are the eigenvalues. The $\chi_{jk} =\int_\mT\int_\mS \{X(s,t)-\mu(s,t)\}\psi_j(s)\phi_k(t)dsdt$ are the marginal projection scores. To be precise, we should first consider expanding $X(s,t)$ in terms of completed versions of the bases of marginal eigenfunctions, but since it can be shown that the scores $\chi_{jk}$ associated with the extra functions needed to complete the bases are 0, the expansion of $X(s,t)$ in \Cref{productFPCA} holds.

The above product functional principal component analysis representation is the same as the Karhunen--Lo\`eve representation if one makes the separable covariance assumption $C(s,t;u,v) = a  C_{1}(s,u) C_{2}(t,v)$, which we call strong separability in contrast to the  weak separability that will be proposed in this paper. However, if strong separability is not assumed, the marginal eigenfunctions no longer carry optimal efficiency guarantees \citep{aston2012evaluating}, and can only be proven to have near-optimality under appropriate assumptions \citep{chen2015modeling}. Moreover, unlike the $Z_l$ in multi-way functional principal component analysis, the scores $\chi_{jk}$ are not guaranteed to be mutually uncorrelated.   

Factorization of the signal into its spatial ($s$) and temporal ($t$) components, justified using a vague notion of spatial-temporal separability, is a common strategy used in many methods in image analysis and multi-way functional data analysis \citep{zhang20052d, lu2006multilinear, huang2009analysis,chen2012modeling, hung2012multilinear, allen2014generalized,chen2015quantifying,chen2015modeling}. Despite their empirical success, the rigorous characterization of this separable feature is still mainly restricted to the scope
of strong separability, i.e., when the covariance $C(s,t;u,v)$ is separable. There is a large amount of literature on strong separability in related fields  \citep{lu2005likelihood,fuentes2006testing,srivastava2009estimation,hoff2011separable,horvath2012inference}. Tests for strong separability in functional data settings have been proposed recently \citep{aston2015tests, constantinou2015testing}. 
 
In this paper, we propose a new concept of weak separability for the process $X$, which can be rigorously tested. We show that under weak separability the eigenfunctions of the full covariance $C(s,t;u,v)$ can be written as tensor products of the marginal eigenfunctions, i.e., $\psi_j\otimes \phi_k$. This means the Karhunen--Lo\`eve representation is the same as the product representation in \Cref{productFPCA}, just as if we had strong separability, and to perform functional principal component analysis we only need to calculate the marginal covariances $C_{\mS}(s,u)$ and $C_{\mT}(t,v)$ instead of the full covariance $C(s,t;u,v)$. The analysis reveals that if $C(s,t;u,v)$ is separable, then the process $X$ is weakly separable, but the converse is not necessarily true. Indeed, weak separability is a much weaker assumption than separable covariance.

We develop a test for weak separability based on the empirical correlations between the estimated scores $\hat \chi_{i,jk}$ and $\hat \chi_{i,j'k'}\ \{i=1,\ldots,n;\ (j,k) \neq (j',k')\}$. 
 Although the $\hat \chi_{i,jk}$ are $n^{1/2}$-consistent estimators of the $\chi_{i,jk}$, the test statistics based on the $\hat \chi_{i,jk}$ have different null distributions from their counterparts using the $\chi_{i,jk}$ due to non-negligible estimation errors.  The proofs involve expansions of the differences between the estimated marginal eigenfunctions and their true values, i.e., $\hat\psi_j - \psi_j$ and $\hat \phi_k - \phi_k$, as well as multi-way tensor products with indices $(j,k,j',k')$. A series of careful derivations are carried out to characterize the asymptotic null distribution of the test statistic, which is found to be a $\chi^2$ type mixture. No Gaussian assumption on $X$ is imposed. We apply the testing procedure to brain imaging data, where frequency and time-based functional connectivity is constructed from source localized magnetoencephalography signals. The test result supports the use of product functional principal component analysis methods and reveals interesting features about brain connectivity over time and frequency.

\rss\section{Weak separability: concepts and properties}\rss
\label{sec:properties}

For $\mS\subseteq\R^{d_1}$ and $\mT\subseteq\R^{d_2}$, we consider the space of square integrable surfaces $L^2(\mS\times \mT)$ with the standard inner product $\langle f, g\rangle =\int_\mT\int_\mS f(s,t)g(s,t) dsdt$ and the corresponding norm  $\|\cdot\|$. 
The data can be viewed as realizations of a random element $X \in L^2(\mS\times \mT)$, which we assume has well defined mean function $\mu$ and covariance operator $C$. We assume the covariance is continuous, and $\mS$ and $\mT$ are compact. Unless otherwise noted, these assumptions are used in all the lemmas and theorems.

For orthonormal bases $f_j\ (j=1,2,\ldots )$ in $L^2(\ms)$ and $g_k\ (k=1,2,\ldots )$ in $L^2(\mt)$, the product functions $f_j(s)g_k(t)\  (j=1,2,\ldots ;\  k=1,2,\ldots )$ form an orthonormal basis of $L^2(\ms\times\mt)$. We can then have
$$X(s,t) = \mu(s,t) + \sum_{j= 1}^{\infty}\sum_{k=1}^{\infty}\tilde \chi_{jk} f_j(s)g_k(t),$$ where $\tilde\chi_{jk} = \int_\mT\int_\mS \{X(s,t)-\mu(s,t)\}f_j(s)g_k(t)dsdt$.  

{\bf Definition of weak separability:} {\it
$X(s,t)$ is weakly separable if there exist orthonormal bases $f_j\ (j=1,2,\ldots )$ and $g_k\ (k=1,2,\ldots )$ such that $\cov(
 \tilde\chi_{jk}, \tilde\chi_{j'k'}) = 0$ for $j\neq j'$ or $k\neq k'$, i.e., the scores $\tilde\chi_{jk}\ (j=1,2,\ldots ;\ k=1,2,\ldots )$ are uncorrelated with each other.
}

In the following, we list several important properties of weak separability, which make this concept attractive in many applications. Detailed proofs are given in the appendix.

\begin{lemma}
  \label{lemma:marginal}
  If $X$ is weakly separable, the pair of bases $f_j\ (j=1,2,\ldots )$ and $g_k\ (k=1,2,\ldots )$ that satisfies weak separability is unique up to a sign, and $f_j(s) \equiv \psi_j(s)$ and $g_k(t) \equiv \phi_k(t)$, where $\psi_j(s)$ and $\phi_k(t)$ are the eigenfunctions of the marginal kernels
  $C_{\mathcal{S}}(s,u)$ and  $C_{\mathcal{T}}(t,v)$ as defined in \Cref{eq:marginalcov}.
Moreover, \begin{equation}
\label{eq:reducedcov}
C(s,t;u,v)=\sum_{j=1}^{\infty}\sum_{k=1}^{\infty}\eta_{jk}\psi_j(s)\phi_k(t)\psi_j(u)\phi_k(v),
\end{equation}
where $\eta_{jk} = \var(\langle X-\mu, \psi_j\otimes \phi_k\rangle)$, and the convergence is absolute and uniform. 
\end{lemma}

Lemma \ref{lemma:marginal} shows that natural basis functions for the factorized
spatial and temporal effects are eigenfunctions of the marginal kernels. Under weak separability, the eigenfunctions of the covariance $C(s,t;u,v)$ can be written as tensor products of the marginal eigenfunctions, i.e., $\psi_j\otimes \phi_k$, which could result in a substantial dimension reduction in applications. Lemma \ref{lemma:marginal} also allows us to test the weak separability assumption (see Section \ref{sec:test}). 


\begin{lemma}
  \label{lemma:strong separability}
  Strong separability, defined as $C(s,t;u,v) = a C_{1}(s,u) C_{2}(t,v)$ with identifiability constraints $\int_{\mathcal{S}}C_{1}(s,s)ds = 1$ and $\int_{\mathcal{T}}C_{2}(t,t)dt = 1$,
  implies weak separability of $X$. And up to a constant scaling, $C_{1}$ and $C_{2}$ are the same as the marginal kernels.
\end{lemma}

Lemma \ref{lemma:strong separability} shows that strong separability is a special case of weak separability, and the following Lemma \ref{lemma:rankone} further illustrates that weak separability is much more flexible than strong separability.

\begin{lemma}
\label{lemma:rankone}
Define the array $V = (\eta_{jk}, \ j=1,2,\ldots ;\ k=1,2,\ldots )$. Strong separability is weak separability with the additional assumption that ${\rm rank}(V) = 1$. Moreover, under strong separability $V = a \Lambda \Gamma^{T}$, where $\Lambda = (\lambda_1,\lambda_2,\ldots)^T$ and $\Gamma = (\gamma_1,\gamma_2,\ldots)^T$ are the eigenvalues of the marginal kernels, and $a = 1/\int_{\mathcal{T}}\int_{\mathcal{S}} C(s,t;s,t)ds dt$ is a normalization constant. 
\end{lemma}

When the covariance $C$ is not strongly separable but the process $X$ is weakly separable, we can show that the covariance function is a sum of $L$ separable components, $C(s,t;u,v) = \sum_{l=1}^L a^l C_{\ms}^l(s,u)C_{\mt}^l(t,v),$ where $L \geq {\rm rank} (V) > 1$ is the nonnegative rank, defined as $${\rm rank}_+(V)=\min\{\ell: V=V_1+\cdots+V_\ell;\ V_i\ge 0,\ {\rm rank}(V_i)=1,\ {\rm for}\ {\rm all}\ i\},$$ where $V_i\ge 0$ means that $V_i$ is entry-wise nonnegative. In applications where one relies on the separable structure of the covariance for ease of computation and interpretation, for example in applications involving the inverse of the covariance, it is not clear whether and how one can modify the concept to work under the weak separability assumption ($L$ additive separable terms). We defer this to future research.

\rss\section{Test of weak separability}\rss
\label{sec:test}

\subsection{Background}
\label{sec:background}

Assume we have a sample of independent and identically distributed smooth processes $X_i(s,t) \sim X(s,t)$, and the marginal projection scores $$\chi_{i, jk} =\int_\mT\int_\mS \{X_i(s,t)-\mu(s,t)\}\psi_j(s)\phi_k(t)dsdt,$$ where $\psi_j(s)$ and $\phi_k(t)$ are the eigenfunctions of the marginal covariances. By the definition of weak separability and Lemma \ref{lemma:marginal}, testing weak separability is the same as testing the covariance structure of the marginal projection scores, i.e., $H_0:$ $\cov(\chi_{jk},\chi_{j'k'}) = 0$ for $j\neq j'$ or $k\neq k'$.

The problem of testing covariance structure is a classic problem in multivariate analysis. Suppose we have independent and identically distributed copies of a $p$-variate random variable, with mean $\mu$ and covariance matrix $\Sigma$, and we want to test the null hypothesis that $\Sigma$ is diagonal.
Under the traditional multivariate setting where $p$ is fixed and does not increase with $n$, likelihood ratio methods can be used to test the diagonality  of $\Sigma$ \citep{anderson1984introduction}. The high-dimensional problem has been studied in the context that $p/n \rightarrow \gamma \in (0, \infty)$ or even for $p$ much larger than $n$ \citep{ledoit2002some,liu2008asymptotic,cai2011limiting, lan2015testing}. If we were to observe the sample values $\chi_{i,jk}$, a sensible test statistic could be based on the off-diagonal terms of the empirical covariance, i.e., $n^{-1/2}\sum_{i=1}^{n}\chi_{i,jk}\chi_{i,j'k'}.$
However, unlike in the traditional covariance testing problem, we do not directly observe the sample values $\chi_{i,jk}$. Instead they are estimated from the sample curves $X_i(s,t)\ (i = 1,\ldots, n)$ as
\begin{equation*}
  \label{eq:estimatescore}
  \hat\chi_{i, jk} =\int_\mT\int_\mS \{X_i(s,t)-\bar X(s,t)\}\hat\psi_j(s)\hat\phi_k(t)dsdt, \end{equation*}
  where $\bar X(s,t) =(1/n) \sum_{i=1}^{n}X_i(s,t)$, and $\hat\psi_j$ and $\hat\phi_k$ are eigenfunctions of the estimated marginal covariances 
 $\hat C_{\mS}(s,u) = (1/n)\sum_{i=1}^{n}\int_{\mathcal{T}} \{X_i(s,t)-\bar X(s,t)\}\{X_i(u,t)-\bar X(u,t)\} dt $  and  $\hat C_{\mT}(t,v) = (1/n)\sum_{i=1}^{n}\int_{\mathcal{S}} \{X_i(s,t)-\bar X(s,t)\}\{X_i(s,v)-\bar X(s,v)\} ds$. In practice, if the data for each subject are observed on arbitrarily dense and equally spaced grid points, and recorded in matrices $X_i\ (i = 1,\ldots, n)$, the above estimators can be simplified as $\hat C_{\mS} = (1/n)\sum_{i=1}^{n} (X_i - \bar X)(X_i -\bar X)^T $,  and  $\hat C_{\mT} = (1/n)\sum_{i=1}^{n} (X_i - \bar X)^T (X_i -\bar X)$. The data cannot immediately be written as matrices if the argument $s$ has dimension greater than 1, but as long as the observations are dense in $\mS$ one can vectorize them along a certain ordering of $s$, compute the marginal covariances, and reorganize back accordingly.

Although we can prove that the $\hat\chi_{i,jk}$ are $n^{1/2}$-consistent estimators of the $\chi_{i,jk}$, test statistics based on $n^{-1/2}\sum_{i=1}^{n}\hat\chi_{i,jk}\hat\chi_{i,j'k'}$ have different null distributions from their counterparts using the $\chi_{i,jk}$, and in the following we derive the asymptotic distribution of the former. 

\subsection{The test statistic and its properties}
\label{sec:testprops}

Let $H$ be a real separable Hilbert space, with inner product $\inner{\cdot}{\cdot}$. Following standard definitions, we denote the space of bounded linear operators on $H$ as $\mathcal{B}(H)$, the space of Hilbert--Schmidt operators on $H$ as $\mathcal{B}_{HS}(H)$, and the space of trace-class operators on $H$ as $\mathcal{B}_{Tr}(H)$. For any trace-class operator $T\in  \mathcal{B}_{Tr}(H)$, we define its trace by $tr(T) = \sum_{i=1,2,\ldots }\inner{Te_i}{e_i}$, where $e_i\ (i=1,2,\ldots )$ is an orthonormal basis of $H$, and it is easy to see that this definition is independent of the choice of basis.

For $H_1$ and $H_2$ two real separable Hilbert spaces, we use $\otimes$ as the standard tensor product, i.e., for $x_1 \in H_1$ and $x_2 \in H_2$, $(x_1\otimes x_2)$ is the operator from $H_2$ to  $H_1$ defined by $(x_1 \otimes x_2) y = \inner{x_2}{y}x_1$ for any $y\in H_2$.
With a bit of abuse of notation, we let $H = H_1 \otimes H_2$ denote the tensor product Hilbert space, which contains all finite sums of $x_1 \otimes x_2$, with inner product $\inner{x_1 \otimes x_2}{y_1 \otimes y_2} = \inner{x_1}{y_1}\inner{x_2}{y_2}$, for $x_1,y_1 \in H_1$ and $x_2, y_2 \in H_2$. For $C_1 \in \mathcal{B}(H_1)$ and $C_2 \in \mathcal{B}(H_2)$, we let $C_1\tilde \otimes C_2$ denote the unique bounded linear operator on $H_1  \otimes H_2$  satisfying $C_1\tilde \otimes C_2 (x_1 \otimes x_2) = C_1 x_1 \otimes C_2 x_2$ for all $x_1 \in H_1, \,  x_2 \in H_2.$


We define  \begin{equation}
  \label{eq:Tn}
	T_n(j,k,j',k') = n^{-1/2}\sum_{i=1}^{n}\hat\chi_{i,jk}\hat\chi_{i,j'k'}\quad  \{(j,k)\neq (j',k')\},
  \end{equation}
and $\mathcal{Z}_n = n^{1/2}(C_n - C)$, where the sample covariance operator is defined as $$C_n =(1/n) \sum_{i=1}^{n}(X_i-\bar X) \otimes (X_i-\bar X).$$
  
The following two conditions are needed for the main theorem below and the corollary following it: 

\noindent Condition 1: For some orthonormal basis $e_j\ (j=1,2,\ldots )$ of $L^2(\mS \times \mT)$, $$\sum_{j=1,2,\ldots}\{\E(\inner{X}{e_j}^4)\}^{1/4}<\infty.$$

\noindent Condition 2: For some integers $P$ and $K$, we have $\delta_{P} = \min_{j=1,\ldots,P}(\lambda_j-\lambda_{j+1}) >0$ and $ \delta_{K} = \min_{k=1,\ldots,K}(\gamma_k- \gamma_{k+1}) >0$.

%

{\bf Remark:}
According to Proposition 5 of \cite{mas2006sufficient}, Condition 1 implies that
$\mathcal{Z}_n$ converges to a Gaussian random element in $\mathcal{B}_{Tr}\{L^2(\mS\times\mT)\}$.

\begin{theorem}
	\label{Th:main}
	Assume Conditions 1 and 2 hold, and that $X$ is weakly separable. For $j,j'=1,\ldots, P$ and $k,k'=1,\ldots, K$ as defined in Condition 2, we have
	
   \noindent (i) for $j\neq j'$ and $k\neq k'$, 
	$$ T_n(j,k,j',k')  =tr\left[\{(\psi_j\otimes\psi_{j'})\tilde\otimes(\phi_k\otimes\phi_{k'})\}\mathcal{Z}_n\right] + o_p(1), $$
	\noindent (ii) for $j=j'$ and $k\neq k'$, 
	\begin{align*}
		T_n(j,k,j,k')  = &
		tr\left[\{(\psi_j\otimes\psi_{j})\tilde\otimes(\phi_k\otimes\phi_{k'})\}\mathcal{Z}_n\right] \\\nonumber
		  +& tr\left([Id_1\tilde\otimes \{\eta_{jk'}(\gamma_{k}-\gamma_{k'})^{-1}\phi_k\otimes\phi_{k'}\}]\mathcal{Z}_n\right) \\\nonumber
		  +&  tr\left([Id_1\tilde\otimes \{\eta_{jk}(\gamma_{k'}-\gamma_{k})^{-1}\phi_{k'}\otimes\phi_{k}\}]\mathcal{Z}_n\right) + o_p(1),
	\end{align*}
	\noindent (iii) for $j\neq j'$ and $k=k'$,
	\begin{align*}
		T_n(j,k,j',k)  =&
		 tr\left[\{(\psi_j\otimes\psi_{j'})\tilde\otimes(\phi_k\otimes\phi_{k})\}\mathcal{Z}_n\right] \\\nonumber
		 + &  tr\left([\{\eta_{jk}(\lambda_{j'}-\lambda_{j})^{-1}\psi_{j'}\otimes\psi_{j}\}\tilde\otimes Id_2]\mathcal{Z}_n\right)\\\nonumber
		 +&  tr\left([\{\eta_{j'k}(\lambda_{j}-\lambda_{j'})^{-1}\psi_{j}\otimes\psi_{j'}\}\tilde\otimes Id_2]\mathcal{Z}_n\right)+o_p(1),
	\end{align*}
	where $Id_1$ and $Id_2$ are identity operators on $L^2(\mS)$ and $L^2(\mT)$, respectively. 
	\end{theorem}
	
{\bf Remark:}	
Since $n^{1/2}tr\left\{(\psi_j\otimes\psi_{j'})\tilde\otimes(\phi_k\otimes\phi_{k'})C\right\}$ is zero under the null hypothesis, the first term in each case of the above theorem is the same as  $n^{1/2}tr\left\{(\psi_j\otimes\psi_{j'})\tilde\otimes(\phi_k\otimes\phi_{k'})C_n\right\} =  n^{-1/2}\sum_{i=1}^{n}\chi_{i,jk}\chi_{i,j'k'}$, i.e., the counterpart of $T_n$ as if we had the true marginal projection scores. The second and third terms, if they exist, are non-negligible estimation errors.

	\begin{corollary}
	\label{corollary:distribution}
 	Assume Conditions 1 and 2 hold, and that $X$ is weakly separable. For different sets of $(j,k,j',k')$, $j,j'=1,\ldots, P$; $k,k'=1,\ldots, K$, satisfying $(j,k)\neq (j',k')$, the $T_n(j,k,j',k')$'s are asymptotically jointly Gaussian with mean zero and covariance structure $\Theta$. The formula for $\Theta$ is given in the proof. 	
 \end{corollary}

%
%
%
\subsection{Tests based on $\chi^2$ type mixtures}
\label{sec:chi2mixture}
	\begin{lemma}
	\label{lemma:covarscores}
		For $j\neq j'$, $\sum_{k=1}^{\infty}\E(\chi_{jk}\chi_{j'k}) = 0$, and for $k\neq k'$, $\sum_{j=1}^{\infty} \E(\chi_{jk}\chi_{jk'}) =0$. This also holds in the empirical version such that for $j\neq j'$, $\sum_{k=1}^{\infty}T_n(j,k,j',k) = 0$, and for $k\neq k'$, $\sum_{j=1}^{\infty}T_n(j,k,j,k') = 0$. 
		\end{lemma}
	The above lemma does not assume weak separability. Recall the fact that principal component scores in traditional functional principal component analysis are uncorrelated. This lemma is a generalized result for the marginal projection scores.
	
Due to this linear relationship between the different terms of $T_n$, the asymptotic covariance $\Theta$ will be degenerate, and thus the statistic we consider is the sum of squares of the terms of $T_n$ without normalizing by the covariance. In practice, for suitably chosen $P_n$ and $K_n$, we use the statistic defined as
	\begin{equation*}
		\label{eq:teststatistic}
	S_{n} =  \sum_{j,j'=1,\ldots, P_n;\ k,k'=1,\ldots,K_n;\ (j,k) < (j',k')} \{T_n(j,k,j',k')\}^2,\end{equation*}
where $(j,k) < (j',k')$ means $(j-1)*K_n + k < (j'-1)*K_n + k'$. 

Take $T_{n}$ to be a long vector of length $m=P_n K_n (P_n K_n -1)/2$ created by stacking all of the $T_{n}(j,k,j',k')\ \{j,j'=1,\ldots, P_n;\ k,k'=1,\ldots, K_n;\ (j,k) < (j',k')\}$. Then by Corollary \ref{corollary:distribution}, $T_{n} \sim N_{m}(0,\Theta)$ under $H_{0}$, where we now take $\Theta$ to be a covariance matrix. Define the spectral decomposition of $\Theta$ as $\Theta = UQ U^{T}$, where $Q$ is diagonal with diagonal entries $\sigma_{1}, \ldots, \sigma_{m}$, which are the eigenvalues of $\Theta$ ordered from largest to smallest, and $U = [u_{1} \ldots \ u_{m}]$, where the $u_{i}$ are orthonormal column vectors. By Lemma \ref{lemma:covarscores}, some of the $\sigma_{i}$ are 0. Since $S_{n} = \Vert  T_{n} \Vert^{2} = \Vert  U^{T}T_{n} \Vert^{2} $ and $U^{T}T_{n} \sim N_{m}(0,Q)$, we can write $S_{n} = \sum_{i=1}^{m} \sigma_{i} A_{i}$ where the $A_{i}$ are independent and identically distributed $\chi_{1}^{2}$, i.e., the null distribution of $S_n$ is a weighted sum of $\chi^2$ distributions, which we call a $\chi^2$ type mixture. 

The Welch--Satterthwaite approximation for a $\chi^2$ type mixture \citep{zhang2013analysis} approximates $S_{n} \sim \beta \chi_{d}^{2}$ and determines $\beta$ and $d$ from matching the first 2 cumulants (the mean and the variance). This results in $\beta = tr(\Theta ^{2})/tr(\Theta)$ and $d = \{tr(\Theta)\}^{2}/tr(\Theta^{2})$. By using a plug-in estimator of $\Theta$, we can approximate the P-value for our test as an upper tail probability of $\beta \chi_{d}^{2}$. When the first $(P_n, K_n)$ terms do not satisfy weak separability, we have $S_n {\rightarrow} \infty$ in probability by noticing that for at least one set of $(j,k,j',k')$, the first term in \Cref{eq:Tnexpansion} in the proof of Theorem \ref{Th:main} is on the order of $n^{1/2}$.


The consistent selection of $(P_n,K_n)$ for hypothesis testing is a challenging problem. The optimal choice of $(P_n,K_n)$ needs to be defined according to the problem at hand and subsequent analysis of interest. Here we focus on the subspace where the subsequent product functional principal component analysis is going to be carried out. A criterion we will use to evaluate a given choice of  $(P_n,K_n)$ is the fraction of variance explained by the first $P_{n}$ and $K_{n}$ components, defined as
\begin{equation*}
 \label{eq:FVE}
\textrm{FVE}({P_{n},K_{n}})={{{1\over{n}}\sum_{i=1}^{n} \sum_{j=1}^{P_{n}} \sum_{k=1}^{K_{n}} \hat{\chi}^{2}_{i,jk}}\over{{1\over{n}}\sum_{i=1}^{n} \sum_{j=1}^{\infty} \sum_{k=1}^{\infty} \hat{\chi}^{2}_{i,jk}}} .
\end{equation*}
This definition can be justified by noting its relation to the normalized mean squared $L^{2}$ loss of the truncated process $\tilde{X}(s,t) = \mu(s,t) + \sum_{j=1}^{P_{n}} \sum_{k=1}^{K_{n}} \chi_{jk} \psi_{j}(s)\phi_{k}(t)$. In particular,
\begin{equation*}
{{\E(\|X-\tilde X\|^2)}\over{\E(\|X - \mu\|^2)}} = 1 - {{\sum_{j=1}^{P_{n}} \sum_{k=1}^{K_{n}}\eta_{jk}}\over{\sum_{j=1}^{\infty} \sum_{k=1}^{\infty} \eta_{jk}}}.
\end{equation*}
The latter term is approximated by our definition of fraction of variance explained. The above equality only relies on the orthogonality of the eigenfunctions, not the weak separability assumptions. Thus, it still makes sense to consider this definition of fraction of variance explained even when $H_{0}$ is not true.

We also define the marginal fractions of variance explained as $\textrm{FVE}_{S}(P_{n})=\sum_{j=1}^{P_{n}} \hat \lambda_{j} / \sum_{j=1}^{\infty} \hat \lambda_{j}$ and $\textrm{FVE}_{T}(K_{n})=\sum_{k=1}^{K_{n}}\hat \gamma_{k} / \sum_{k=1}^{\infty}\hat \gamma_{k}$, where the $\hat{\lambda}_{j}$ are the eigenvalues of $\hat C_{\ms}$ and the $\hat{\gamma}_{k}$ are the eigenvalues of $\hat C_{\mt}$. In practice the infinite sums in the denominators of $\textrm{FVE}({P_{n},K_{n}})$, $\textrm{FVE}_{S}(P_{n})$, and  $\textrm{FVE}_{T}(K_{n})$ will have to be replaced with the largest number of terms that can reasonably be considered nonzero.
  
Noting that $\sum_{j=1}^{\infty}\E(\chi_{jk}^2) = \gamma_k$, $\sum_{k=1}^{\infty}\E(\chi_{jk}^2) = \lambda_j$ and $\sum_{j=1}^{\infty}\sum_{k=1}^{\infty}\E(\chi_{jk}^2) = \sum_{j=1}^{\infty} \lambda_j = \sum_{k=1}^{\infty} \gamma_k $, we have $$\textrm{FVE}(P_n,K_n) \gtrsim  \textrm{FVE}_{\mS}(P_n)+\textrm{FVE}_{\mT}(K_n) - 1,$$ subject to estimation error (to see, for example, that $\sum_{k=1}^{\infty}\E(\chi_{jk}^2) = \lambda_j$, take $j=j'$ in the proof of Lemma \ref{lemma:covarscores}, with no need to assume weak separability). Therefore, we propose the following fraction of variance explained procedure: First choose $P_{n}$ and $K_{n}$ such that the marginal fractions of variance explained are at least 90\%. If this choice results in $\textrm{FVE}(P_{n},K_{n}) \geq 90\%$, use these values of $P_{n}$ and $K_{n}$. If not, use the values of $P_{n}$ and $K_{n}$ that have marginal fractions of variance explained at least 95\%, in which case $\textrm{FVE}(P_{n},K_{n})$ is expected to be above 90\%.  

\subsection{Bootstrap approximation}
\label{sec:bootstrap}

As an alternative to asymptotic approximation, we can also consider a bootstrap approach to approximate the distribution of the test statistic.  Theorem \ref{Th:main} provides theoretical support for the use of the following empirical bootstrap procedure  \citep{van1996weak}. Our simulations show that the asymptotic approximation based on the $\chi^2$ type mixture has very satisfactory performance. We still present the bootstrap approximation here since it is generally applicable to  similar tests where the asymptotic null distributions do not have closed form. 


At each step, draw a random sample from the data $X_{1},\ldots,X_{n}$ with replacement. Denote this sample as $X^{*}_{1},\ldots,X^{*}_{n}$. Let 
\begin{equation*}
\label{eq:estscoresboot}
  \hat\chi_{i, jk}^{*} =\int_\mT\int_\mS \{X^{*}_i(s,t)-\bar X^{*}(s,t)\}\hat\psi^{*}_j(s)\hat\phi^{*}_k(t)dsdt, \end{equation*}
where $\bar X^{*}$ is the sample mean of the $X^{*}_i$, and the $\hat\psi^{*}_j$ and $\hat\phi^{*}_k$ are the eigenfunctions of the estimated marginal covariances of the $X^{*}_i$. The signs of the $\hat\psi^{*}_j$ and $\hat\phi^{*}_k$ are chosen to minimize $\Vert \hat{\psi}_{j}^{*}-\hat{\psi}_{j}  \Vert$ and $\Vert \hat{\phi}_{k}^{*}-\hat{\phi}_{k}  \Vert$, respectively.  Let 
\begin{equation*}
\label{eq:teststatboot}
	T^{*}_n(j,k,j',k') = n^{-1/2}\sum_{i=1}^{n}\hat\chi^{*}_{i,jk}\hat\chi^{*}_{i,j'k'}.
  \end{equation*}
 The empirical bootstrap test statistic is calculated as 
 \begin{equation*}
	S_{n}^{*} =  \sum_{j,j'=1,\ldots, P_n;\ k,k'=1,\ldots,K_n;\ (j,k) < (j',k')} \{T_n^{*}(j,k,j',k')-T_n(j,k,j',k')\}^2	.\end{equation*}
 This procedure is repeated $B$ times, and the P-value is approximated as the proportion of bootstrap test statistics $S_{n}^{*}$ that are larger than the test statistic $S_{n}$. 

Theorem 3.9.13 in \cite{van1996weak} can be used to prove the validity of the bootstrap procedure, i.e., the conditional random laws (given data) of $S_n^*$ are asymptotically consistent almost surely for estimating the laws of $S_n$, under the null hypothesis.
By Theorem \ref{Th:main}, we have that under the null hypothesis, $T_n$ can be written as $\Phi_P'\{n^{1/2}(\mathcal{P}_n-P)\}+ o(1)$ and $T_n^*-T_n$ can be written as $\Phi_P'\{n^{1/2}(\mathcal{P}_n^*-\mathcal{P}_n)\}+o(1)$, where $\Phi_P'$ is a linear continuous mapping that depends on the three different cases in Theorem \ref{Th:main}. Thus, Theorem 3.9.13 applies.

Other than the above non-studentized empirical bootstrap based on $S_n$, we have also considered a bootstrap procedure based on a marginally studentized test statistic, in which we divide each term in $S_n$ by its corresponding estimated variance $\hat{\theta}(j,k,j',k')$ (which is the plug-in estimate of ${\theta}(j,k,j',k')$, the diagonal entry of $\Theta$ corresponding to the asymptotic variance of $T_n(j,k,j',k')$). However, we have found this procedure is much more time consuming, and requires substantially higher sample size to achieve high power, in comparison to the non-studentized empirical bootstrap method. 
This is not unexpected, since the form of ${\theta}(j,k,j',k')$ is very complicated and plug-in estimation adds extra variability. Therefore, we do not recommend the marginally studentized empirical bootstrap method.

\rss\section{Numerical study}\rss
\label{sec:simulations}

In this section, we conduct numerical experiments to check the size and power of the proposed test for weak separability. 
Following the notation in Section \ref{sec:test}, we are basically testing if the $\cov(\chi_{jk}, \chi_{j'k'})$ are all zero for $(j,k)\neq (j',k')$. 
We consider two different choices for the joint distribution of the ${\chi}_{jk}$. The first is the multivariate normal and the second is the multivariate $t$ distribution. The diagonal values of $\Sigma$, the covariance matrix of the ${\chi}_{jk}$, are determined by the matrix $V= \{\var(\chi_{jk}), \, j,k=1,\ldots,8\}$. We consider two different choices for $V$, which we denote as $V_{1}$ and $V_{2}$ (specified later). Under $H_0$ (when all of the off-diagonal values of $\Sigma$ are 0), $V_{1}$ corresponds to a strongly separable covariance structure, while $V_{2}$ corresponds to a weakly separable structure that is not strongly separable.
To study power, for a given choice of $V_{1}$ or $V_{2}$, we take $\textrm{cov}(\chi_{i,12},\chi_{i,21})$ to be the largest positive value such that $\Sigma$ is positive definite, and we also consider half of this value. Alternatively, we let 3 off-diagonal terms, $\textrm{cov}(\chi_{i,12},\chi_{i,21})$,  $\textrm{cov}(\chi_{i,11},\chi_{i,22})$, and $\textrm{cov}(\chi_{i,13},\chi_{i,31})$, take their largest positive values such that $\Sigma$ is positive definite. 
 
Empirical rejection rates at the .05 significance level 
from 200 simulations runs for $n=50, 100, 500$ are shown in Tables \ref{tab:chi2weakseplongV1} through \ref{tab:nostudweakseplongV2}.  The fraction of variance explained method described in Section \ref{sec:chi2mixture} ends up with $P_{n}=3$ and $K_{n}=2$ in most trials, and we show results with $(P_n, K_n)$ chosen by this procedure, as well as directly setting $(P_n, K_n)=(2,2)$, $(3,3)$, or $(4,4)$ for all trials.  We see that both the $\chi^2$ type mixture approximation and the empirical bootstrap procedure are able to control the type I error under all scenarios and achieve very good power as $n$ 
or the signal increase, although the empirical bootstrap is slightly less powerful for small $n$. Even when the chosen nonzero off-diagonal covariance terms are set to their maximum values, the other off-diagonal covariance terms of $\Sigma$ are zero, and so the signal is moderate. The test procedures are slightly less powerful in the multivariate $t$ case for small $n$, as the asymptotics likely come into play more quickly for the normal data. The rejection rates are in general stable across different choices of $(P_n,K_n)$; although $(P_n, K_n) = (2,2)$ seems to have higher power in some cases, the power stabilizes to a reasonable value for larger $(P_n,K_n)$.

        \begin {table} [hbt]
\caption {Rejection rates for the $\chi^2$ type mixture weak separability test procedure, using $V_1$ and choosing $(P_n,K_n)$ with the fraction of variance explained procedure (FVE) or as $(2,2)$, $(3,3)$, or $(4,4)$
} \label{tab:chi2weakseplongV1}
   \begin{tabular} {  p{3.8cm}  p{0.9cm}   p{0.9cm} p{0.9cm}  p{1.3cm}   p{0.9cm}  p{0.9cm}  p{0.9cm}   p{0.9cm} }
       Scenario & \multicolumn{4}{l}{\hspace{1.7cm}Normal}  & \multicolumn{4}{c}{Multivariate $t$}   \\
    $n=50$     & FVE & (2,2) & (3,3) & (4,4)  & FVE & (2,2) & (3,3) & (4,4)    \\ 
  $H_0$ &    0.055  &  0.020 &   0.020 &   0.040 &  0.025 &  0.020  &  0.005  &  0.045 \\
    $\textrm{cov}(\chi_{12},\chi_{21})=0.065$ &  0.715 &   0.785  &  0.740 &   0.755  &  0.440 &   0.445 &   0.395 &   0.370 \\
  $\textrm{cov}(\chi_{12},\chi_{21})=0.13$ &    1.000  &  1.000  &  1.000  &  1.000  &  0.935  &  0.965  &  0.940  &  0.940 \\ 
  3 nonzero terms   &    1.000  &  1.000  &  1.000  &  1.000  &  0.960  &  0.990  &  0.990  &  0.965 \\
    $n=100$ & & & & & & & & \\
  $H_0$ &    0.035 &   0.075 &   0.045  &  0.050  &  0.025  &  0.040  & 0.020  &  0.010 \\
    $\textrm{cov}(\chi_{12},\chi_{21})=0.065$ &    0.985 &   0.985 &   0.985  &  0.985  &  0.800  &  0.810  &  0.785  &  0.710 \\
  $\textrm{cov}(\chi_{12},\chi_{21})=0.13$ &    1.000 &   1.000  &  1.000  &  1.000  &  1.000 &   0.990  &  0.990  &  0.990 \\
  3 nonzero terms   &    1.000 &   1.000  &  1.000  &  1.000 &   1.000 &   0.990  &  1.000   & 1.000 \\
   $n=500$ & & & & & & & & \\ 
  $H_0$ &    0.060  &  0.060  &  0.040  &  0.055  &  0.020  &  0.045  &  0.020  &  0.045 \\ 
    $\textrm{cov}(\chi_{12},\chi_{21})=0.065$ &    1.000  &  1.000  &  1.000  &  1.000  &  0.995  &  0.995  &  1.000  &  0.990 \\
  $\textrm{cov}(\chi_{12},\chi_{21})=0.13$ &    1.000  &  1.000  &  1.000  &  1.000  &  1.000  &  1.000  &  1.000  &  1.000 \\
  3 nonzero terms   &    1.000  &  1.000  &  1.000  &  1.000  &  1.000   & 1.000  &  1.000  &  1.000 
   \end{tabular} 
\end {table}    
    
        \begin {table} [hbt]
\caption {Rejection rates for the $\chi^2$ type mixture weak separability test procedure, using $V_2$ and choosing $(P_n,K_n)$ with the fraction of variance explained procedure (FVE) or as $(2,2)$, $(3,3)$, or $(4,4)$
} \label{tab:chi2weakseplongV2}
   \begin{tabular} {  p{3.8cm}  p{0.9cm}   p{0.9cm} p{0.9cm}  p{1.3cm}   p{0.9cm}  p{0.9cm}  p{0.9cm}   p{0.9cm} }
       Scenario & \multicolumn{4}{l}{\hspace{1.7cm}Normal}  & \multicolumn{4}{c}{Multivariate $t$}   \\
    $n=50$     & FVE & (2,2) & (3,3) & (4,4)  & FVE & (2,2) & (3,3) & (4,4)    \\ 
  $H_0$ &    0.030  &  0.025  &  0.030  &  0.025  &  0.015  &  0.030  &  0.015  &  0.005 \\
  $\textrm{cov}(\chi_{12},\chi_{21})=0.055$ &    0.515  &  0.845  &  0.440  &  0.465  &  0.305  &  0.555  &  0.225  &  0.205 \\
    $\textrm{cov}(\chi_{12},\chi_{21})=0.11$ &    0.995 &  0.995  &  0.995  &  0.990  &  0.850 &   0.955 &   0.825  &  0.770 \\
    3 nonzero terms   &    1.000 &   1.000  &  1.000 &   1.000  &  0.965  &  0.985  &  0.950  &  0.970 \\
    $n=100$ & & & & & & & & \\
  $H_0$ &    0.045  &  0.055 &   0.035  &  0.040  &  0.010  &  0.050  &  0.040  &  0.020 \\
  $\textrm{cov}(\chi_{12},\chi_{21})=0.055$ &    0.920  &  0.990  &  0.930  &  0.920  &  0.625  &  0.900  &  0.605  &  0.500 \\
  $\textrm{cov}(\chi_{12},\chi_{21})=0.11$ &    1.000  &  1.000  &  1.000  &  1.000  &  0.990  &  1.000  &  0.965   & 0.955 \\
    3 nonzero terms   &    1.000  &  1.000 &   1.000  &  1.000  &  0.995  &  1.000  &  1.000  &  0.980 \\
   $n=500$ & & & & & & & & \\ 
  $H_0$ &    0.045 &   0.065  &  0.025  &  0.040  &  0.025  &  0.065  &  0.050  &  0.035 \\
  $\textrm{cov}(\chi_{12},\chi_{21})=0.055$ &    1.000  &  1.000  &  1.000  &  1.000  &  0.970  &  1.000  &  0.995  &  0.995 \\
  $\textrm{cov}(\chi_{12},\chi_{21})=0.11$ &    1.000  &  1.000  &  1.000  &  1.000  &  1.000  &  1.000 &   0.990  &  0.995 \\
    3 nonzero terms   &    1.000  &  1.000  &  1.000  &  1.000  &  1.000  &  1.000  &  1.000  &  1.000 
   \end{tabular} 
\end {table}

        \begin {table} [hbt]
\caption {Rejection rates for the non-studentized empirical bootstrap weak separability test procedure, using $V_1$ and choosing $(P_n,K_n)$ with the fraction of variance explained procedure (FVE) or as $(2,2)$, $(3,3)$, or $(4,4)$
} \label{tab:nostudweakseplongV1}
   \begin{tabular} {  p{3.8cm}  p{0.9cm}   p{0.9cm} p{0.9cm}  p{1.3cm}   p{0.9cm}  p{0.9cm}  p{0.9cm}   p{0.9cm} }
       Scenario & \multicolumn{4}{l}{\hspace{1.7cm}Normal}  & \multicolumn{4}{c}{Multivariate $t$}   \\
           $n=50$     & FVE & (2,2) & (3,3) & (4,4)  & FVE & (2,2) & (3,3) & (4,4)    \\ 
$H_0$ &    0.035 &   0.025  &  0.035 &   0.025  &  0.010  &  0.010 &   0.005 &   0.005 \\
$\textrm{cov}(\chi_{12},\chi_{21})=0.065$ &    0.670 &   0.680 &   0.650  &  0.640  &  0.350  &  0.375 &   0.330 &   0.300 \\
$\textrm{cov}(\chi_{12},\chi_{21})=0.13$ &    1.000 &   1.000 &   1.000  &  1.000  &  0.890  &  0.900 &   0.885  &  0.880 \\
  3 nonzero terms   &    1.000 &   1.000  &  1.000  &  1.000  &  0.920  &  0.920 &   0.920  &  0.915 \\
    $n=100$ & & & & & & & & \\
$H_0$ &    0.070  &  0.060 &   0.060 &  0.060  &  0.015  &  0.010  &  0.015 &   0.015 \\
$\textrm{cov}(\chi_{12},\chi_{21})=0.065$ &    0.990 &   0.995 &   0.985 &   0.985  &  0.735 &   0.785 &   0.700  &  0.680 \\
$\textrm{cov}(\chi_{12},\chi_{21})=0.13$ &    1.000 &   1.000  &  1.000  &  1.000  &  0.970 &   0.970 &   0.965  &  0.960 \\
  3 nonzero terms   &    1.000  & 1.000  &  1.000  &  1.000  &  1.000  &  1.000  &  1.000  &  1.000 \\
   $n=500$ & & & & & & & & \\     
$H_0$ &    0.065  &  0.050  &  0.060  &  0.060 &   0.055  &  0.050  &  0.055 &   0.055 \\
$\textrm{cov}(\chi_{12},\chi_{21})=0.065$ &    1.000  &  1.000  &  1.000  &  1.000 &   0.995  &  0.995 &   0.995 &   0.995 \\
$\textrm{cov}(\chi_{12},\chi_{21})=0.13$ &    1.000 &   1.000 &   1.000 &   1.000  &  1.000  &  1.000 &   1.000 &   1.000 \\
  3 nonzero terms   &    1.000  &  1.000  &  1.000 &   1.000  &  1.000  &  1.000  &  1.000  &  1.000
       \end{tabular} 
\end {table}   
    
        \begin {table} [hbt]
\caption {Rejection rates for the non-studentized empirical bootstrap weak separability test procedure, using $V_2$ and choosing $(P_n,K_n)$ with the fraction of variance explained procedure (FVE) or as $(2,2)$, $(3,3)$, or $(4,4)$
} \label{tab:nostudweakseplongV2}
   \begin{tabular} {  p{3.8cm}  p{0.9cm}   p{0.9cm} p{0.9cm}  p{1.3cm}   p{0.9cm}  p{0.9cm}  p{0.9cm}   p{0.9cm} }
       Scenario & \multicolumn{4}{l}{\hspace{1.7cm}Normal}  & \multicolumn{4}{c}{Multivariate $t$}   \\
           $n=50$     & FVE & (2,2) & (3,3) & (4,4)  & FVE & (2,2) & (3,3) & (4,4)    \\ 
$H_0$ &    0.065  &  0.045  &  0.050 &   0.045 &   0.015 &   0.020 &   0.010  &  0.010 \\
$\textrm{cov}(\chi_{12},\chi_{21})=0.055$ &    0.420  &  0.785  &  0.390  &  0.380 &   0.220  &  0.395  &  0.160  &  0.150 \\
$\textrm{cov}(\chi_{12},\chi_{21})=0.11$ &    0.970  &  1.000  &  0.965  &  0.965 &   0.730 &   0.865 &   0.690  &  0.675 \\
  3 nonzero terms   &    1.000  &  1.000  &  1.000 &   1.000  &  0.895  &  0.925 &   0.885  &  0.885 \\
    $n=100$ & & & & & & & & \\
$H_0$ &    0.025  &  0.010 &   0.020 &   0.020 &   0.035 &   0.035  &  0.030  &  0.020 \\
 $\textrm{cov}(\chi_{12},\chi_{21})=0.055$ &   0.950  &  1.000 &   0.955 &   0.955 &   0.585 &   0.855  &  0.575  &  0.515 \\
$\textrm{cov}(\chi_{12},\chi_{21})=0.11$ &    1.000  &  1.000 &   1.000 &   1.000 &   0.980 &   0.995  &  0.980  &  0.975 \\
  3 nonzero terms   &    1.000 &   1.000 &    1.000 &    1.000 &   0.985 &   0.990 &   0.985  &  0.985 \\
   $n=500$ & & & & & & & & \\     
$H_0$ &    0.030  &  0.065  &  0.030 &   0.030 &   0.010 &   0.035  &  0.005   &      0 \\
 $\textrm{cov}(\chi_{12},\chi_{21})=0.055$ &   1.000 &   1.000  &  1.000 &   1.000 &   1.000 &   1.000  &  1.000  &  1.000 \\
$\textrm{cov}(\chi_{12},\chi_{21})=0.11$ &    1.000 &   1.000  &  1.000  &  1.000  &  1.000  &  1.000  &  1.000  &  1.000 \\
  3 nonzero terms   &    1.000 &   1.000 &   1.000  &  1.000  &  1.000 &   1.000  &  1.000  &  1.000
       \end{tabular} 
\end {table}

Details of the simulation settings are as follows:
We generate independent samples of data $X_{i}(s,t) = \sum_{j=1}^{8} \sum_{k=1}^{8} \chi_{i,jk} \psi_{j}(s)\phi_{k}(t)\ (i=1,\ldots,n)$, where the scores $\chi_{i,jk}$ are mean 0 random variables that we generate directly. We let $s$ and $t$ take values from 0 to 1 on an evenly spaced grid of 20 points. For the $\psi_{j}$ we use the functions $\psi_{j} (s)= -2^{1/2}\cos\{\pi(n+1)s\}$ for $j$ odd and $\psi_{j} (s)= 2^{1/2}\sin(\pi ns)$ for $j$ even.  
We define the $\phi_{k}$ by taking the first 3 B-spline  functions produced by Matlab's spcol function using order 4 with knots at 0, 0.5, and 1, combining these with the first 5 $\psi_{j}$ as defined above, and orthonormalizing using Gram--Schmidt. 

Let ${\chi}_{i}$ be the vector of $\chi_{i,jk}$ for $j,k=1,\ldots,8$. 
We simulate each ${\chi}_{i}$ independently from either $N(0,\Sigma)$ or  the multivariate $t$ distribution. In the latter case, we first simulate a vector $x$ of length 64 from $N(0,\Sigma)$. One standard definition of a multivariate $t$ vector is $x/(u/v)^{1/2}$, where $u$ is a chi-square random variable with $v$ degrees of freedom that is independent of $x$. However, we use $x/\{u/(v-2)\}^{1/2}$ as our multivariate $t$ vector so that its covariance matrix is $\Sigma$. We take $v=6$ in our simulations. For each of 200 trials, we simulate data $X_{i}(s,t)\ (i=1,\ldots,n)$ in the manner described above, estimate the marginal projection scores, calculate the test statistic, and obtain P-values from the test procedures as described in Section \ref{sec:test}, using $B= 1000$ for the bootstrap procedure. 

We choose $V_1$ and $V_2$ to both give $\lambda_j = \exp\{1.2(9-j)\}/\left\{\sum_{j'=1}^{8} \exp(1.2j')\right\}\ (j=1,\ldots,8)$ and  $\gamma_k = \exp\{1.6(9-k)\}/\left\{\sum_{k'=1}^{8} \exp(1.6k')\right\}\ (k=1,\ldots,8)$ as the eigenvalues of the marginal covariances $C_{\mS}$ and $C_{\mT}$. $V_1$ is defined as the rank 1 matrix computed from the outer product of the vectors of $\lambda_j$ and $\gamma_k$, while $V_2$ is a rank 2 matrix with first 2 rows multiples of each other and rows 3 through 8 multiples of each other.

\section{Application to brain connectivity studies}
\label{sec:application}

Brain imaging analysis is an area where functional data increasingly arise. An important goal in brain imaging studies is to analyze functional connectivity between different regions of the brain. We focus on magnetoencephalography (MEG), which measures neuronal activity by recording magnetic fields generated within the brain. We use MEG data collected by the Human Connectome Project, a study that has compiled a large amount of high quality multi-modal neural data, much of which is freely accessible at 
https://db.humanconnectome.org
 \citep{van2013wu,wu20171200}. 
We will focus on the motor task data, particularly the trials where subjects moved their right hand. The signal for each trial is recorded from -1.2 to 1.2 seconds in intervals of about 2 ms, where time 0 corresponds to the start of the motion.  
In the preprocessed sensor-level data (see \cite{wu20171200} for preprocessing details), there are 61 subjects with motor data, and the subjects have an average of 75.38 trials. Our connectivity analysis will focus on two regions of interest: the left primary motor cortex and the right inferior parietal lobule. These regions of interest are spatially separated, likely activated during the task, and potentially functionally connected.  

As the MEG sensors are distant from the brain, directly using their signals to represent regions of interest can lead to spurious connectivity measurements. This is due to the volume conduction/field spread problem, in which each sensor picks up the activity of several sources, as well as the common input problem, in which a common source provides input to a pair of signals that do not directly interact \citep{larson2013adding, bastos2015tutorial}. For these reasons, we will use source reconstruction to estimate the signals arising from the cortical surface. Source reconstruction is common in MEG analysis, but it is an inverse problem on which constraints must be placed to obtain a unique solution \citep{pizzella2014magnetoencephalography}. The source reconstruction method we use is minimum norm estimation as implemented in the Matlab package FieldTrip \citep{lin2004spectral,oostenveld2010fieldtrip}. 

MEG signals are inherently oscillatory, and synchronization at certain frequency ranges of the activity in different regions has been shown to be related to tasks performed by the brain \citep{pizzella2014magnetoencephalography}. To study how frequency-based coupling between regions of interest changes over the course of a task, we calculate the time-frequency representations of their signals based on Morlet wavelets, using FieldTrip's ft\_freqanalysis function. The time-frequency representation of a signal is its representation at time $t$ and frequency $s$ as a complex number $A(s,t)e^{i B(s,t) }$, where $A(s,t)$ is the amplitude and $B(s,t)$ is the phase. 

Given time-frequency representations $A_{1,k}(s,t)e^{i B_{1,k}(s,t) }$ and $A_{2,k}(s,t)e^{i B_{2,k}(s,t) }$ for two signals recorded in trial $k$, $k=1,\ldots,n_T$, we use the phase locking value \citep{lachaux1999measuring} to measure their connectivity, calculated as 
\begin{equation*}
\rm{PLV}(s,t)=(1/n_T) \vert \sum_{k=1}^{n_T} e^{i\{B_{1,k}(s,t)-B_{2,k}(s,t)\}} \vert.
\end{equation*}
The phase locking value takes values from 0 to 1, with 1 indicating complete phase synchrony over trials and 0 indicating no phase synchrony. The phase locking value, like many connectivity measures, is based on an analogue of the cross-correlation function called the coherence, but the phase locking value disregards the amplitudes and considers only the magnitude of the average of the phase differences as unit vectors in the complex plane. The phase locking value has gained popularity due to the belief that phase differences reveal more about functional connectivity than changes in amplitude \citep{lachaux1999measuring, aydore2013note, bastos2015tutorial}.

Because we calculate the time-frequency representation using wider time windows for lower frequencies, we are limited in how low of frequencies we can consider, and our preliminary results for power show a lack of activity above 50 Hz. 
Thus, we calculate the time-frequency representation from 8 to 50 Hz, corresponding to the alpha to gamma low frequency bands. In each trial, the motion usually lasts no longer than about 0.75 seconds. The signal at a time period shortly before time 0 is of interest, as it can represent brain activity when subjects have received the movement cue but have not yet reacted to it. However, the trials are not disjoint, so the signal at times further before 0 overlaps with the signal from the previous trial. Thus, in our analysis we will consider times for each trial on the range of -0.25 to 0.75 seconds. Calculating phase locking value between the two source-reconstructed signals corresponding to our regions of interest, the data we analyze is $\textrm{PLV}_i(s,t)$, where $i=1,\ldots, 61$; $8 \leq s \leq 50$; and $-0.25 \leq t \leq 0.75$.

\begin{figure}[!htbp]
  \noindent\makebox[\textwidth]{ 
   \includegraphics[width=1.2\textwidth]{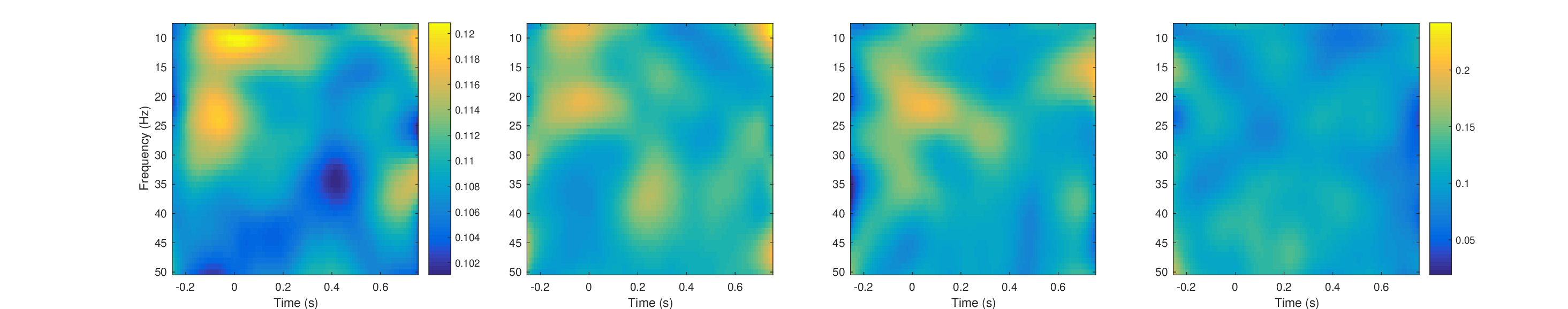}}
  \caption{Plots of the average source-level phase locking value (left), and the source-level phase locking value for 3 subjects. The latter plots use the same color scale, which is different than that used for the average.} 
  \label{Fi:indivplvsource}
\end{figure}

Figure \ref{Fi:indivplvsource} shows the average of the phase locking value matrices over all subjects, as well as slightly smoothed phase locking value matrices for 3 randomly selected subjects. The level of activity seems to vary between subjects. The average phase locking value displays higher synchrony near the beginning of the movement (time 0) in the alpha and beta bands, and the individual subjects' plots also show higher values near time 0. However, the average has small values overall, which indicates high variability between subjects, and points to the need to study covariance structure and modes of variation.

For the phase locking value data described above, using the fraction of variance explained procedure described in Section \ref{sec:chi2mixture}, we choose the number of components to be $P_n=7$ and $K_n=7$. We apply the weak separability test using both the $\chi^2$ type mixture approximation (P-value = 0.5293) and the empirical bootstrap (P-value = 0.9260). The weak separability test does not reject the null hypothesis of weak separability, which supports the use of functional principal component analysis based on products of marginal eigenfunctions. We also apply the strong separability test of \cite{aston2015tests} via their R package covsep€ \citep{tavakoli2016covsep}. The resulting P-values are $1.198\times 10^{-4}$ for their chi-square approximation and 0.08 for their non-studentized empirical bootstrap method. 

\begin{figure}[!htbp]
   \noindent\makebox[\textwidth]{ 
   \includegraphics[width=0.6\textwidth]{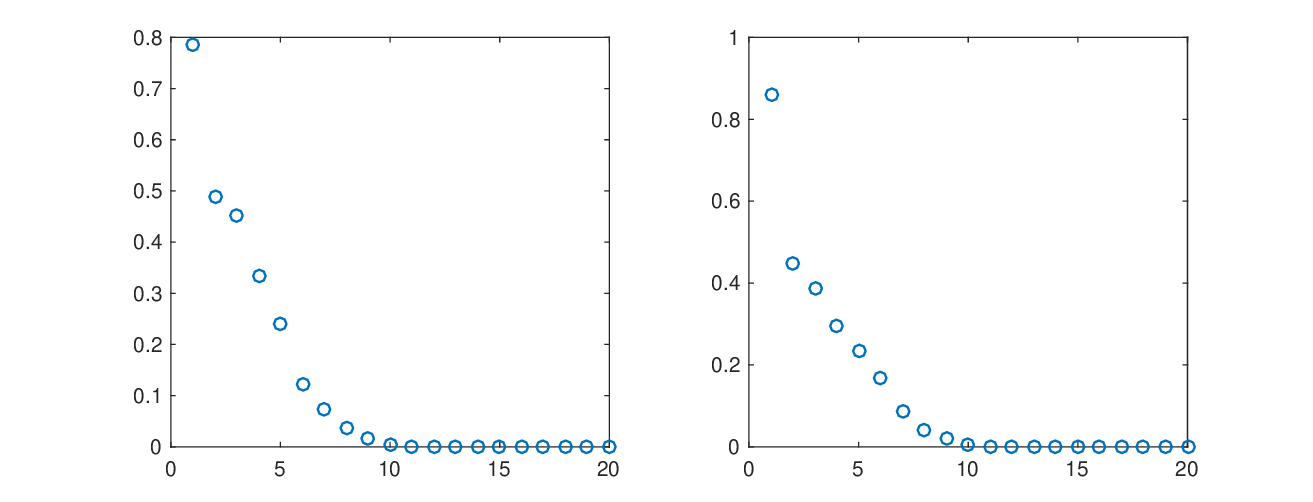}}
   \caption{Plots of the first 20 estimated marginal eigenvalues $\hat{\lambda}_j$ (left, for frequency) and $\hat{\gamma}_k$ (right, for time).} 
   \label{Fi:lambdasourcemotor}
 \end{figure}

Product functional principal component analysis represents the data $\textrm{PLV}_i(s,t)$ with products $\psi_j(s)\phi_k(t)$ of the marginal eigenfunctions, where  $\psi_j(s)$ represents the frequency component and $\phi_k(t)$ represents the time component. The 3  estimated eigenfunction products that account for the most variance are $\hat{\psi}_1(s)\hat{\phi}_1(t)$, $\hat{\psi}_2(s)\hat{\phi}_1(t)$, and $\hat{\psi}_3(s)\hat{\phi}_1(t)$. The variance explained by $\hat{\psi}_1(s)\hat{\phi}_1(t)$ is by far the largest. The estimated marginal eigenvalues $\hat{\lambda}_j$ and $\hat{\gamma}_k$ are plotted in Figure \ref{Fi:lambdasourcemotor}. We see the first eigenvalue dominates the others, and there is also a slight drop between the second two $\hat{\lambda}_j$ and the rest, reflecting the fact that $\hat{\psi}_2(s)\hat{\phi}_1(t)$ and $\hat{\psi}_3(s)\hat{\phi}_1(t)$ are the products that explain the second and third highest amounts of variance, respectively.

\begin{figure}[!htbp]
   \noindent\makebox[\textwidth]{ 
   \includegraphics[width=1.3\textwidth]{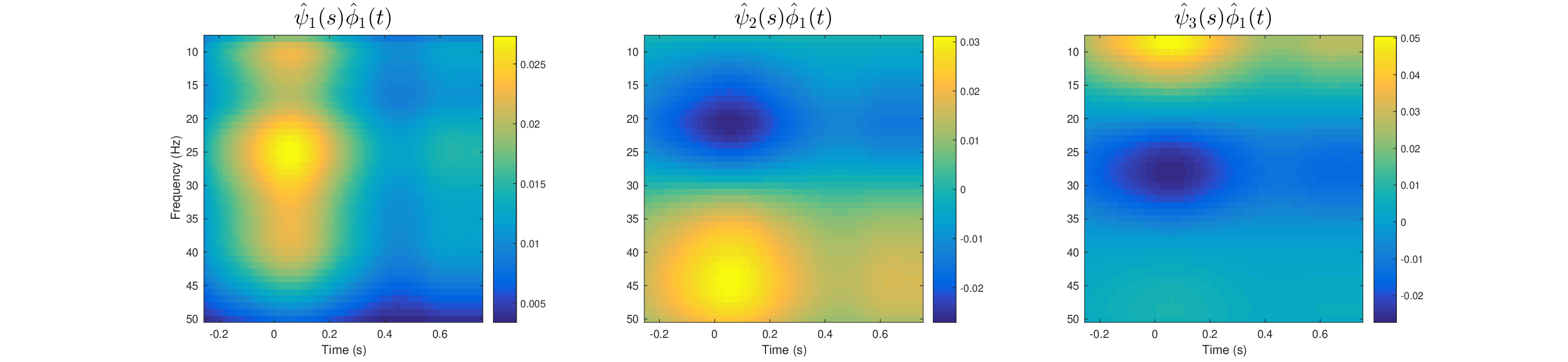}}
   \caption{Plots of the products of the estimated eigenfunctions that explain the most variance.} 
   \label{Fi:prodsourcemotor}
 \end{figure}

\begin{figure}[!htbp]
	   \noindent\makebox[\textwidth]{ 
	   \includegraphics[width=0.6\textwidth]{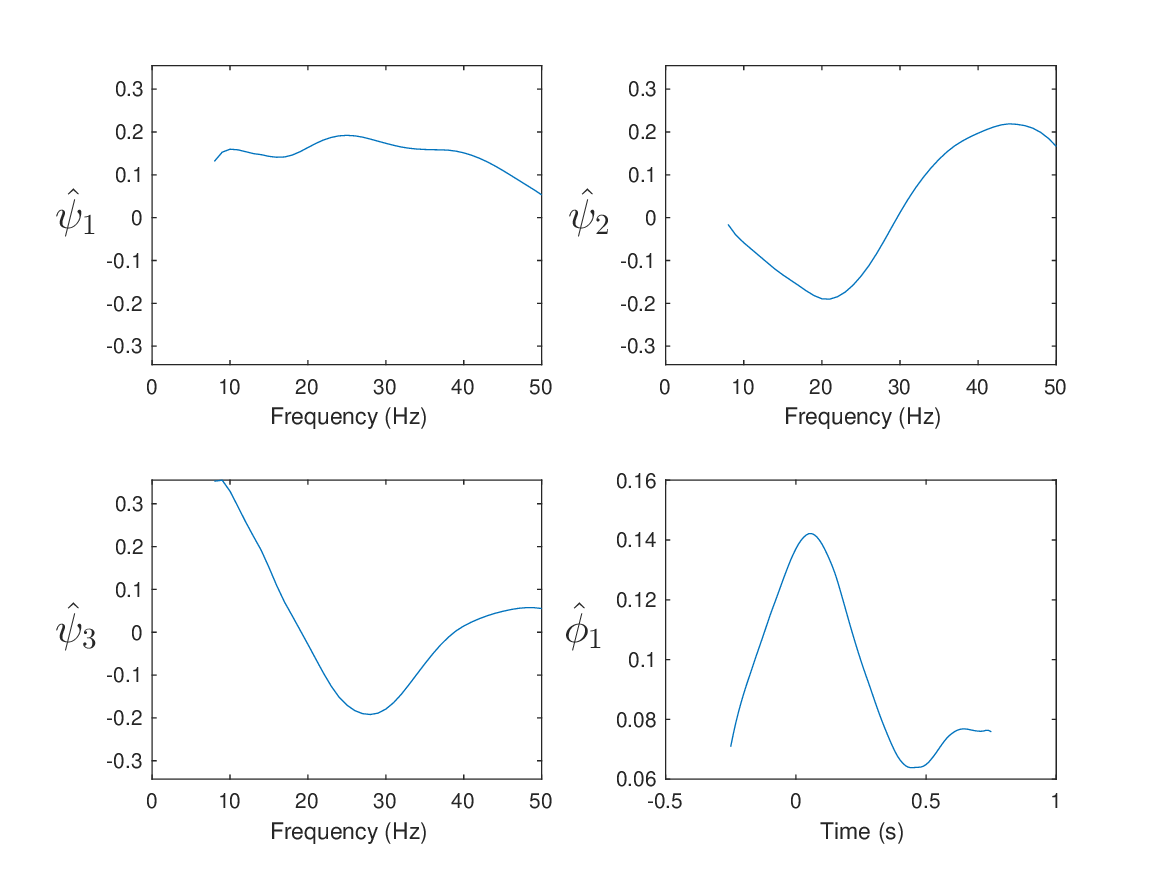}}
	   \caption{Plots of the estimated eigenfunctions $\hat{\psi}_j(s)$ and $\hat{\phi}_k(t)$ whose products explain the most variance.} 
	   \label{Fi:psisourcemotor}
	 \end{figure}

The product functions $\hat{\psi}_1(s)\hat{\phi}_1(t)$, $\hat{\psi}_2(s)\hat{\phi}_1(t)$, and $\hat{\psi}_3(s)\hat{\phi}_1(t)$ are plotted in Figure \ref{Fi:prodsourcemotor}. These products capture modes of variation mainly around -0.2 to 0.2 s, from when the subject receives the cue to move to when they just start moving. This variation can be seen more clearly in the first temporal eigenfunction $\hat{\phi}_1(t)$ (shown on the bottom right of Figure \ref{Fi:psisourcemotor}, which plots the individual marginal eigenfunctions), which peaks slightly after 0 s. $\hat{\psi}_1(s)\hat{\phi}_1(t)$ shows that, within this time range, subjects generally vary in synchrony from the alpha band to the beginning of the gamma low band, peaking within the beta band around 20--30 Hz. $\hat{\psi}_2(s)\hat{\phi}_1(t)$ shows a contrast between the beta low band and gamma low band. That is, subjects with higher $\chi_{21}$ values have lower synchrony in the beta low band and higher synchrony in the gamma low band. $\hat{\psi}_3(s)\hat{\phi}_1(t)$ shows a contrast between the alpha band and the beta high band. 
  

\section{Discussion}
Much of the benefit of using product functional principal component analysis under weak separability is related to ease of  interpretation and computation; by representing the eigenfunctions as tensor products of the marginal eigenfunctions, one consumes far fewer degrees of freedom and only needs to compute the marginal covariances instead of the full covariance. When the weak separable assumption does not hold, the product functional principal component analysis scores are correlated, and one expects to have to use more terms in product functional principal component analysis than in conventional functional principal component analysis to explain the same amount of variance. Product functional principal component analysis can still be used under the alternative as a dimension reduction approach, but one needs to be aware of the above issues. We believe that the notion of weak separability will inspire new methodological developments for multi-way functional data analysis, such as multi-way regularization on marginal components \citep{chen2015localized}. 

Although our data example has $s$ and $t$ in $\R$, our test works for scenarios where $d_1>1$ or $d_2>1$. For example, when modeling brain imaging data $X_i(s,t)$ observed on a dense grid of voxels or dipoles over the cortical surface, in which $s \in \R^2$ or $s \in \R^3$, one can first vectorize along $s$ by ordering the dipoles from 1 to $M$, compute the marginal covariances, perform the hypothesis test of weak separability, and reorganize back to the  space $\mS$ for interpretation and visualization. However, in scenarios where data are only very sparsely observed on the domain, where individual-subject smoothing is not appropriate, the problem is much more challenging and beyond the scope of this paper. We will possibly pursue it in future work.

\section*{Acknowledgement}
Data were provided in part by the Human Connectome Project, WU-Minn Consortium (Principal Investigators: David Van Essen and Kamil Ugurbil; 1U54MH091657) funded by the 16 NIH Institutes and Centers that support the NIH Blueprint for Neuroscience Research; and by the McDonnell Center for Systems Neuroscience at Washington University.


\rss\section{Appendix: proofs}\rss
\noindent{\bf Proof of Lemma \ref{lemma:marginal}}

Let $f_{j}\ (j =1,2,\ldots )$ and $g_{k}\ (k =1,2,\ldots )$ be a pair of bases that satisfies weak separability. For $(j,k)\neq (j',k')$, we have 
$\langle C f_j\otimes g_k, f_{j'}\otimes g_{k'}\rangle = E \left(\langle X-\mu,f_j\otimes g_k\rangle\langle X-\mu, f_{j'}\otimes g_{k'}\rangle\right) 
 = 0.$
Since the covariance operator $C$ is diagonalized under the orthonormal basis $f_j\otimes g_k\ (j=1,2,\ldots ;\ k=1,2,\ldots )$, by Mercer's theorem, $$C(s,t;u,v)= \sum_{j=1}^{\infty} \sum_{k=1}^{\infty} \eta_{jk} f_{j}(s)g_{k}(t)f_{j}(u)g_{k}(v),$$ where
$\eta_{jk} = \langle C f_j\otimes g_k, f_{j}\otimes g_{k}\rangle = \var \left( \langle X-\mu,f_j\otimes g_k\rangle \right)$, and the convergence is absolute and uniform.  

The marginal kernel can then be written as
\begin{align*}
C_{\mathcal{S}}(s,u)= &\int_{\mathcal{T}} \sum_{j=1}^{\infty} \sum_{k=1}^{\infty}\eta_{jk}f_{j}(s)g_{k}(t)f_{j}(u)g_{k}(t) dt\\\nonumber
 = & \sum_{j=1}^{\infty} \left(\sum_{k=1}^{\infty}\eta_{jk} \right) f_{j}(s)f_{j}(u).
\end{align*}
The exchange of the integral and sums is allowed by the Fubini--Tonelli theorem, by noticing that 
\begin{align*}
&\int_{\mathcal{T}} \sum_{j=1}^{\infty} \sum_{k=1}^{\infty} |\eta_{jk}f_{j}(s)g_{k}(t)f_{j}(u)g_{k}(t)| dt \\ \nonumber
 \leq & \int_{\mathcal{T}} \left\{\sum_{j=1}^{\infty} \sum_{k=1}^{\infty} \eta_{jk} f^2_{j}(s)g^2_{k}(t)\right\}^{1/2}\left\{\sum_{j=1}^{\infty} \sum_{k=1}^{\infty} \eta_{jk} f^2_{j}(u)g^2_{k}(t)\right\}^{1/2} dt \\ \nonumber
 = &  \int_{\mathcal{T}} C(s,t; s,t)^{1/2}C(u,t;u,t)^{1/2} dt\\ \nonumber
 \leq & \int_{\mathcal{T}} \sup_{s,t}|C(s,t,s,t)| dt \leq \infty,
\end{align*}
where we use the Cauchy--Schwarz inequality. 

Thus, we see that the $f_{j}$ are eigenfunctions of $C_{\mathcal{S}}$ with eigenvalues $\lambda_{j} =\sum_{k=1}^{\infty} \eta_{jk}$. An analogous computation shows that the $g_{k}$ are eigenfunctions of $C_{\mathcal{T}}$ with eigenvalues $\gamma_{k} =\sum_{j=1}^{\infty} \eta_{jk}$. 

\noindent{\bf Proof of Lemma \ref{lemma:strong separability}}

With strong separability, we have $C(s,t;u,v)=a{C}_{1}(s,u){C}_{2}(t,v)$. From the definition of $C_{\mathcal{S}}$, we have
$$C_{\mathcal{S}}(s,u)= \int_{\mathcal{T}} C(s,t;u,t)dt = a {C}_{1}(s,u) \int_{\mathcal{T}} {C}_{2}(t,t) dt = a {C}_{1}(s,u).$$
An analogous argument shows ${C}_{\mathcal{T}}(t,v)=a {C}_{2}(t,v)$. Note that $a=\int_{\mathcal{T}}\int_{\mathcal{S}} C(s,t;s,t) ds dt$.
If we use the marginal eigenfunctions $\psi_j$ and $\phi_k$ as the bases, it is easy to show that when $(j,k) \neq (j',k')$,
$\cov(\chi_{jk},\chi_{j'k'})  = \int_{\mathcal{T},\mathcal{S},\mathcal{T},\mathcal{S}} C(s,t;u,v) \psi_{j}(s)\phi_{k}(t)\psi_{j'}(u)\phi_{k'}(v) ds dt du dv = 0$. Thus, we have weak separability.

\noindent{\bf Proof of Lemma \ref{lemma:rankone}}

When $V$ is of rank 1, $V$ can be written $V=WZ^{T}$, where $W$ and $Z$ are column vectors with entries $(w_{1},w_{2},\ldots)$ and $(z_{1},z_{2},\ldots)$, respectively. Thus, $\eta_{jk}=w_{j}z_{k}$, and under weak separability, \Cref{eq:reducedcov} can be written 
\begin{align*}
C(s,t;u,v) 
=& \sum_{j=1}^{\infty} \sum_{k=1}^{\infty}  w_{j}z_{k} \psi_{j}(s)\psi_{j}(u)\phi_{k}(t)\phi_{k}(v) \\\nonumber
=& \left\{ \sum_{j=1}^{\infty} w_{j} \psi_{j}(s)\psi_{j}(u) \right\} \left\{\sum_{k=1}^{\infty} z_{k} \phi_{k}(t)\phi_{k}(v) \right\}.
\end{align*}
The above can be normalized to fit the definition of strong separability in Lemma  \ref{lemma:strong separability}. 

Under strong separability, from the proof of Lemma 2 we have $$C(s,t;u,v) = \frac{1}{\int_{\mathcal{T}}\int_{\mathcal{S}} C(s,t;s,t) ds dt} C_{\mathcal{S}}(s,u) C_{\mathcal{T}}(t,v),$$ so $\eta_{jk} = \{1/ \int_{\mathcal{T}}\int_{\mathcal{S}} C(s,t;s,t) ds dt\} \lambda_{j}\gamma_{k}$, and then $V = \{1/ \int_{\mathcal{T}}\int_{\mathcal{S}} C(s,t;s,t) ds dt\} \Lambda \Gamma^{T}$.  

\noindent{\bf Proof of Theorem \ref{Th:main}}

For $H_1$ and $H_2$ two real separable Hilbert spaces, we further define the partial trace with respect to $H_1$ as the unique bounded linear operator $tr_1: \mathcal{B}_{Tr}(H_1\otimes H_2) \rightarrow \mathcal{B}_{Tr}(H_2)$ satisfying $tr_{1}(C_1 \tilde \otimes C_2) = tr(C_1)C_2$ for all $C_1 \in \mathcal{B}_{Tr}(H_1)$, $C_2 \in \mathcal{B}_{Tr}(H_2)$. The partial trace with respect to $H_2$ is defined symmetrically and denoted by $tr_{2}$.
With the notation of partial trace, we can see that $ C_{\mT} = tr_1( C)$ and $C_{\mS} = tr_2 (C)$.
 The estimated marginal covariance operators can also be written as $\hat {C}_{\mS} = tr_2({ C}_n)$ and  $\hat { C}_{\mT} = tr_1({C}_n)$. We use these equalities in proofs but not in computation. In practice, the estimated marginal covariances are calculated without having to calculate $C_n$. 
 
We use similar notation and conditions as used by \cite{aston2015tests}. However, to derive the asymptotic distribution of their test statistic for strong separability, they focus on deriving the asymptotic distribution of the difference between the sample covariance operator and its strong separable approximation. Then by projecting on the estimated marginal eigenfunctions, they check the requirement for strong separability that $\eta_{jk} = a\lambda_j\gamma_k$. They do not need further results on the estimation errors of the marginal eigenfunctions and random scores besides that they are consistent. By contrast, our proofs involve the expansion of $\hat\psi_j - \psi_j$ and $\hat \phi_k - \phi_k$, and four-way tensor products with indices $(j,k,j',k')$.

From Condition 1 in Section \ref{sec:testprops} and the remark following it,
	$\mathcal{Z}_n = n^{1/2}(C_n-C)$ converges to a Gaussian random element in $\mathcal{B}_{Tr}\{L^2(\mS\times\mT)\}$ with mean 0 and covariance structure 
	$\Sigma_C = \E[\{(X-\mu)\otimes (X-\mu) - C\}\tilde \otimes \{(X-\mu)\otimes (X-\mu) - C\}].$
	
For $T_n$ as defined in \Cref{eq:Tn},	
	$$T_n(j,k,j',k') = n^{1/2}\inner{C_n(\hat\psi_j\otimes\hat\phi_k)}{\hat\psi_{j'}\otimes\hat\phi_{k'}} = n^{1/2}tr\{(\hat\psi_j\otimes\hat\psi_{j'})\tilde\otimes(\hat\phi_k\otimes\hat\phi_{k'})C_n\}.$$
	Using (5.1.8) in \cite{hsing2015theoretical}, we have $$(\hat\psi_j-\psi_j) = \mathcal{M}_j(\hat C_\mathcal{S} - C_\mathcal{S})\psi_j + o_p(\hat\psi_j-\psi_j)  ,$$
	where $\mathcal{M}_j = \sum_{m\neq j}(\lambda_j-\lambda_m)^{-1}\psi_m\otimes\psi_m \in \mathcal{B}_{Tr}(\mathcal{S})$ and $\lambda_j$ is the $j$th eigenvalue of $C_\mathcal{S}$. Analogously,  
	$$(\hat\phi_k-\phi_k) =  \mathcal{M'}_k(\hat C_\mathcal{T} - C_\mathcal{T})\phi_k + o_p(\hat\phi_k-\phi_k),$$
		where $\mathcal{M'}_k =\sum_{m\neq k}(\gamma_k-\gamma_m)^{-1}\phi_m\otimes\phi_m  \in \mathcal{B}_{Tr}(\mathcal{T})$ and $\gamma_k$ is the $k$th eigenvalue of $C_\mathcal{T}$. Here, Condition 2 is used to guarantee that $\mathcal{M}_j $ and $\mathcal{M'}_k$ exist for $j=1,\ldots,P$ and $k=1,\ldots, K$. 
		
Using $\hat C_\mathcal{S} - C_\mathcal{S} = tr_{2}(C_{n}-C)$ and $\hat C_\mathcal{T} - C_\mathcal{T} = tr_{1}(C_{n}-C)$, we can write $T_n(j,k,j',k')$ as 
\begin{align}
\label{eq:Tnexpansion}
T_n(j,k,j',k') 
=&n^{1/2}tr\left\{(\psi_j\otimes\psi_{j'})\tilde\otimes(\phi_k\otimes\phi_{k'})C\right\}\\\nonumber
& +n^{1/2}tr\left\{(\psi_j\otimes\psi_{j'})\tilde\otimes(\phi_k\otimes\phi_{k'})(C_n-C)\right\}\\\nonumber
& + n^{1/2}tr\left((\psi_j\otimes\psi_{j'})\tilde\otimes[\phi_k\otimes\{\mathcal{M'}_{k'} tr_1(C_n - C)\phi_{k'}\}]C\right)\\\nonumber
&  + n^{1/2}tr\left((\psi_j\otimes\psi_{j'})\tilde\otimes[\{\mathcal{M'}_k tr_1(C_n - C)\phi_{k}\}\otimes \phi_{k'}]C\right)\\\nonumber
&  + n^{1/2}tr\left([\psi_j\otimes\{\mathcal{M}_{j'} tr_2(C_n - C)\psi_{j'}\}]\tilde\otimes(\phi_k\otimes\phi_{k'})C\right)\\\nonumber
&  + n^{1/2}tr\left([\{\mathcal{M}_j tr_2(C_n - C)\psi_{j}\}\otimes\psi_{j'}]\tilde\otimes(\phi_k\otimes\phi_{k'})C\right)\\\nonumber
& + o_p(1).
\end{align}
The first term in the above equation is zero under $H_0$, since under $H_0$  we have the representation $C(s,t,u,v) = \sum_{j=1}^{\infty} \sum_{k=1}^{\infty}\eta_{jk}\psi_j(s)\psi_j(u)\phi_k(t)\phi_k(v),$ where $\eta_{jk}=\var(\chi_{jk})$. Also, by Proposition C.1 in \cite{aston2015tests}, we have that $tr\{A tr_1(T)\}  = tr\{(Id_1\tilde\otimes A) T\}$, where $Id_1$ is an identity operator on $\mathcal{S}$, $A \in \mathcal{B}(\mathcal{T})$, and $T \in \mathcal{B}_{Tr}(\mathcal{S}\times \mathcal{T})$. An analogous identity holds for $tr_2(T)$. 
 Using these facts, we give a simplified form of $T_n(j,k,j',k')$ under $H_0$ for 3 cases:

\noindent (Case i) $j\neq j'$ and $k\neq k'$:
$$ T_n(j,k,j',k')  = tr\left[\{(\psi_j\otimes\psi_{j'})\tilde\otimes(\phi_k\otimes\phi_{k'})\}\mathcal{Z}_n\right] + o_p(1). $$
\noindent (Case ii) $j=j'$ and $k\neq k'$:
\begin{align*}
	T_n(j,k,j',k')  = & tr\left[\{(\psi_j\otimes\psi_{j'})\tilde\otimes(\phi_k\otimes\phi_{k'})\}\mathcal{Z}_n\right] \\\nonumber
	  + & tr\left([Id_1\tilde\otimes \{\eta_{jk'}(\phi_k\otimes\phi_{k'})\mathcal{M'}_k\}]\mathcal{Z}_n\right) \\\nonumber
	  + & tr\left([Id_1\tilde\otimes \{\eta_{jk}(\phi_{k'}\otimes\phi_{k})\mathcal{M'}_{k'}\}]\mathcal{Z}_n\right) + o_p(1).
\end{align*}
\noindent (Case iii) $j\neq j'$ and $k=k'$:
\begin{align*}
	T_n(j,k,j',k')  = & tr\left[\{(\psi_j\otimes\psi_{j'})\tilde\otimes(\phi_k\otimes\phi_{k'})\}\mathcal{Z}_n\right] \\\nonumber
	 + & tr\left([\{\eta_{jk}(\psi_{j'}\otimes\psi_{j})\mathcal{M}_{j'}\}\tilde\otimes Id_2]\mathcal{Z}_n\right) \\\nonumber
	  + & tr\left([\{\eta_{j'k}(\psi_j\otimes\psi_{j'})\mathcal{M}_{j}\}\tilde\otimes Id_2]\mathcal{Z}_n\right) + o_p(1).
\end{align*}
In each of the above cases, two or more of the terms in \Cref{eq:Tnexpansion} end up being zero due to the orthogonality of the eigenfunctions. The latter 2 cases can be simplified to get the result in the statement of the theorem by noting that
$\eta_{jk'}(\phi_k\otimes\phi_{k'})\mathcal{M'}_k = \eta_{jk'}(\gamma_{k}-\gamma_{k'})^{-1}\phi_k\otimes\phi_{k'}$ and $\eta_{jk}(\psi_{j'}\otimes\psi_{j})\mathcal{M}_{j'} = \eta_{jk}(\lambda_{j'}-\lambda_{j})^{-1}\psi_{j'}\otimes\psi_{j}$.

\noindent{\bf Proof of Corollary \ref{corollary:distribution}}

From Theorem \ref{Th:main}, we can see that all the terms of $T_n(j,k,j',k')$ can be written in the form 
$tr\{(A_1\tilde\otimes A_2) \mathcal{Z}_n\}$ for some $A_1 \in \mathcal{B}(\mathcal{S})$ and $A_2 \in \mathcal{B}(\mathcal{T})$. Since $\mathcal{Z}_n$ converges to a Gaussian random element and  $tr\{(A_1\tilde\otimes A_2) \mathcal{Z}_n\}$ is a continuous linear mapping, the
$T_n(j,k,j',k')$ are asymptotically jointly Gaussian for different sets of $(j,k,j',k')$. 
Let $\Theta$ be the covariance structure of the asymptotic joint distribution of the $T_n(j,k,j',k')$, and define $\mathcal{Z}$ to be a Gaussian random element with the limiting distribution of $\mathcal{Z}_n$. By the continuous mapping theorem, 
$\Theta$ can be calculated from terms of the form
\begin{equation}\label{eq:EtraceAB}\E[tr\{(A_1\tilde\otimes A_2) \mathcal{Z}\}tr\{(B_1\tilde\otimes B_2) \mathcal{Z}\}] = 
tr\left\{(A_1\tilde\otimes A_2)\widetilde\bigotimes(B_1\tilde\otimes B_2)\Sigma_C\right\},\end{equation}
where $\Sigma_C$ is defined as in the proof of Theorem \ref{Th:main}.

Recall the Karhunen--Lo\`eve expansion of the process $$X(s,t) = \mu(s,t) + \sum_{j=1}^{\infty} \sum_{k=1}^{\infty}\chi_{jk}\psi_j(s)\phi_{k}(t).$$
We define $u_{ij} = \psi_i\otimes \psi_j \in \mathcal{B}_{HS}(\mathcal{S})$, $v_{ij} = \phi_i\otimes \phi_j \in \mathcal{B}_{HS}(\mathcal{T})$, $\beta_{i,i',j,j',k,k',l,l'} = \E(\chi_{ii'}\chi_{jj'}\chi_{kk'}\chi_{ll'})$, and $\eta_{ii'} = \E(\chi_{ii'}^2)$. 
With weak separability, 
we have 
\begin{align*}
	& tr\left\{(A_1\tilde\otimes A_2)\widetilde\bigotimes(B_1\tilde\otimes B_2)\Sigma_C\right\}\\ \nonumber
= &  \sum_{i,i',j,j',k,k',l,l'}\beta_{i,i',j,j',k,k',l,l'}tr(A_1u_{ij})tr(A_2v_{i'j'})tr(B_1u_{kl})tr(B_2v_{k'l'})\\ \nonumber
& - \sum_{i,i',j,j'}\eta_{ii'}\eta_{jj'}tr(A_1u_{ii})tr(B_1u_{jj})tr(A_2v_{i'i'})tr(B_2v_{j'j'}). 
\end{align*}
Each of the trace terms in the above equation can be evaluated using the identities $tr(Id_{1} u_{ij}) = I(i=j)$, $tr(Id_{2} v_{i'j'}) = I(i'=j')$, $tr\{(\psi_{j_{1}}\otimes\psi_{j_{1}'}) u_{ij}\} = I(i=j_{1})I(j=j_{1}')$, and $tr\{(\phi_{k_{1}}\otimes\phi_{k_{1}'}) v_{i'j'}\} = I(i'=k_{1})I(j'=k_{1}')$. From these identities and the possible forms of $A_1$, $A_2$, $B_1$, and $B_2$ given in Theorem \ref{Th:main}, it follows that the second sum is always 0. The first sum can be simplified by considering 9 cases, as follows:

\noindent (Case 1) $A_1=a_{1}\psi_{j_{1}}\otimes\psi_{j_{1}'}$, $A_2 = a_{2}\phi_{k_{1}}\otimes\phi_{k_{1}'}$, $B_1=b_{1}\psi_{j_{2}}\otimes\psi_{j_{2}'}$, $B_2 = b_{2}\phi_{k_{2}}\otimes\phi_{k_{2}'}$:
 $$tr\left\{(A_1\tilde\otimes A_2)\widetilde\bigotimes(B_1\tilde\otimes B_2)\Sigma_C\right\} = a_1 a_2 b_1 b_2 \beta_{j_{1},k_{1},j_{1}',k_{1}',j_{2},k_{2},j_{2}',k_{2}'}.$$ 
\noindent (Case 2) $A_1=Id_1$, $A_2 = a_{2}\phi_{k_{1}}\otimes\phi_{k_{1}'}$, $B_1=b_{1}\psi_{j_{2}}\otimes\psi_{j_{2}'}$, $B_2 = b_{2}\phi_{k_{2}}\otimes\phi_{k_{2}'}$:
 $$tr\left\{(A_1\tilde\otimes A_2)\widetilde\bigotimes(B_1\tilde\otimes B_2)\Sigma_C\right\} =  a_2 b_1 b_2 \sum_{i=1}^{\infty} \beta_{i,k_{1},i,k_{1}',j_{2},k_{2},j_{2}',k_{2}'}.$$ 
\noindent (Case 3) $A_1=a_{1}\psi_{j_{1}}\otimes\psi_{j_{1}'}$, $A_2 = Id_2$, $B_1=b_{1}\psi_{j_{2}}\otimes\psi_{j_{2}'}$, $B_2 = b_{2}\phi_{k_{2}}\otimes\phi_{k_{2}'}$:
 $$tr\left\{(A_1\tilde\otimes A_2)\widetilde\bigotimes(B_1\tilde\otimes B_2)\Sigma_C\right\} = a_1 b_1 b_2 \sum_{i'=1}^{\infty} \beta_{j_{1},i',j_{1}',i',j_{2},k_{2},j_{2}',k_{2}'}.$$  
\noindent (Case 4) $A_1=a_{1}\psi_{j_{1}}\otimes\psi_{j_{1}'}$, $A_2 = a_{2}\phi_{k_{1}}\otimes\phi_{k_{1}'}$, $B_1=Id_1$, $B_2 = b_{2}\phi_{k_{2}}\otimes\phi_{k_{2}'}$:
 $$tr\left\{(A_1\tilde\otimes A_2)\widetilde\bigotimes(B_1\tilde\otimes B_2)\Sigma_C\right\} = a_1 a_2  b_2 \sum_{k=1}^{\infty} \beta_{j_{1},k_{1},j_{1}',k_{1}',k,k_{2},k,k_{2}'}.$$ 
\noindent (Case 5) $A_1=a_{1}\psi_{j_{1}}\otimes\psi_{j_{1}'}$, $A_2 = a_{2}\phi_{k_{1}}\otimes\phi_{k_{1}'}$, $B_1=b_{1}\psi_{j_{2}}\otimes\psi_{j_{2}'}$, $B_2 = Id_2$:
 $$tr\left\{(A_1\tilde\otimes A_2)\widetilde\bigotimes(B_1\tilde\otimes B_2)\Sigma_C\right\} = a_1 a_2 b_1  \sum_{k'=1}^{\infty}\beta_{j_{1},k_{1},j_{1}',k_{1}',j_{2},k',j_{2}',k'}.$$
\noindent (Case 6) $A_1=Id_1$, $A_2 = a_{2}\phi_{k_{1}}\otimes\phi_{k_{1}'}$, $B_1=Id_1$, $B_2 = b_{2}\phi_{k_{2}}\otimes\phi_{k_{2}'}$:
 $$tr\left\{(A_1\tilde\otimes A_2)\widetilde\bigotimes(B_1\tilde\otimes B_2)\Sigma_C\right\} =  a_2  b_2 \sum_{i=1}^{\infty}\sum_{k=1}^{\infty}\beta_{i,k_{1},i,k_{1}',k,k_{2},k,k_{2}'}.$$    
\noindent (Case 7) $A_1=Id_1$, $A_2 = a_{2}\phi_{k_{1}}\otimes\phi_{k_{1}'}$, $B_1=b_{1}\psi_{j_{2}}\otimes\psi_{j_{2}'}$, $B_2 = Id_2$:
 $$tr\left\{(A_1\tilde\otimes A_2)\widetilde\bigotimes(B_1\tilde\otimes B_2)\Sigma_C\right\} =  a_2 b_1 \sum_{i=1}^{\infty}\sum_{k'=1}^{\infty} \beta_{i,k_{1},i,k_{1}',j_{2},k',j_{2}',k'}.$$  
\noindent (Case 8) $A_1=a_{1}\psi_{j_{1}}\otimes\psi_{j_{1}'}$, $A_2 = Id_2$, $B_1=Id_1$, $B_2 = b_{2}\phi_{k_{2}}\otimes\phi_{k_{2}'}$:
 $$tr\left\{(A_1\tilde\otimes A_2)\widetilde\bigotimes(B_1\tilde\otimes B_2)\Sigma_C\right\} = a_1 b_2 \sum_{i'=1}^{\infty}\sum_{k=1}^{\infty}\beta_{j_{1},i',j_{1}',i',k,k_{2},k,k_{2}'}.$$
\noindent (Case 9) $A_1=a_{1}\psi_{j_{1}}\otimes\psi_{j_{1}'}$, $A_2 = Id_2$, $B_1=b_{1}\psi_{j_{2}}\otimes\psi_{j_{2}'}$, $B_2 = Id_2$:
 $$tr\left\{(A_1\tilde\otimes A_2)\widetilde\bigotimes(B_1\tilde\otimes B_2)\Sigma_C\right\} =  a_1 b_1 \sum_{i'=1}^{\infty}\sum_{k'=1}^{\infty}\beta_{j_{1},i',j_{1}',i',j_{2},k',j_{2}',k'}.$$   
In the above, $a_1$, $a_2$, $b_1$, and $b_2$ are scalar constants. Using the above, all the terms in $\Theta$ can be obtained from straightforward but tedious calculations.

To illustrate the calculation of $\Theta(j,k,j',k',l,m,l',m')$, the term in $\Theta$ corresponding to the asymptotic covariance of $T_n(j,k,j',k')$ and $T_n(l,m,l',m')$, we consider as an example the case where $j\neq j'$, $k\neq k'$, $l\neq l'$, and $m\neq m'$. Here,
\begin{align*}
&  \Theta(j,k,j',k',l,m,l',m')  \\ \nonumber
 \stackrel{by\ Thm.\ \ref{Th:main}\ (i)\ }{=}   &  \E\left(tr\left[\{(\psi_j\otimes\psi_{j'})\tilde\otimes(\phi_k\otimes\phi_{k'})\}\mathcal{Z}\right]tr\left[\{(\psi_l\otimes\psi_{l'})\tilde\otimes(\phi_m\otimes\phi_{m'})\}\mathcal{Z}\right]\right)\\ \nonumber 
 \stackrel{by\ Eq.\ (\ref{eq:EtraceAB})\ }{=}   &   tr\left[\{(\psi_j\otimes\psi_{j'})\tilde\otimes(\phi_k\otimes\phi_{k'})\}\widetilde\bigotimes\{(\psi_l\otimes\psi_{l'})\tilde\otimes(\phi_m\otimes\phi_{m'})\}\Sigma_C\right]\\ \nonumber 
 \stackrel{by\ Case\ 1\ }{=}    &   \beta_{j,k,j',k',l,m,l',m'} = \E(\chi_{jk}\chi_{j'k'}\chi_{lm}\chi_{l'm'}),
\end{align*}
 where we have used  $A_1=\psi_j\otimes\psi_{j'}$, $A_2=\phi_k\otimes\phi_{k'}$, $B_1=\psi_l\otimes\psi_{l'}$, and $B_2=\phi_m\otimes\phi_{m'}$.

\noindent{\bf Proof of Lemma \ref{lemma:covarscores}}

Let $X_N(s,t) = \mu(s,t) + \sum_{j=1}^N\sum_{k=1}^{N}\chi_{jk}\psi_j(s)\phi_k(t)$, and let $C_N$ denote the covariance structure of $X_N$. Thus, $$C_N(s,t;u,v) = \sum_{j=1}^{N}\sum_{j'=1}^{N}\sum_{k=1}^{N}\sum_{k'=1}^{N}\cov(\chi_{jk},\chi_{j'k'})\psi_j(s)\psi_{j'}(u)\phi_k(t)\phi_{k'}(v).$$
It is easy to show that $C_N$ converges to $C$ in Hilbert--Schmidt norm. Let $C_{\mS,N} = tr_2(C_N)$, which converges to $C_{\mS}$ because $tr_2$ is continuous and linear. 
We know that $\langle C_{\mS}\psi_j, \psi_{j'}\rangle = 0$ for $j\neq j'$.  Therefore, for any $\epsilon > 0$, we can find an $N$ such that
  $|\langle C_{\mS, N}\psi_j, \psi_{j'}\rangle| < \epsilon$.
	
	By definition,
	\begin{align*}
	&\langle C_{\mS, N}\psi_j, \psi_{j'}\rangle =\int_{\mS}\int_{\mS} \left\{ \int_{\mT} C_{N}(s,t;u,t)dt \right\} \psi_j(s)\psi_{j'}(u)dsdu \\ \nonumber
	= & \int_{\mS}\int_{\mS} \int_{\mT}  \sum_{l=1}^{N}\sum_{l'=1}^{N}\sum_{k=1}^{N}\sum_{k'=1}^{N}\cov(\chi_{lk},\chi_{l'k'})\psi_l(s)\psi_{l'}(u)\phi_k(t)\phi_{k'}(t) \psi_j(s)\psi_{j'}(u) dtdsdu \\ \nonumber
	= & \sum_{k=1}^{N}\cov(\chi_{jk},\chi_{j'k}) 
	\end{align*}
Therefore, $\lim_N \sum_{k=1}^{N}\cov(\chi_{jk},\chi_{j'k}) = 0$, i.e., $\sum_{k=1}^{\infty}\cov(\chi_{jk},\chi_{j'k}) = 0$ for $j\neq j'$.
 
The same argument holds for the empirical version. Analogous calculations can be done for $k\neq k'$ to show that $\sum_{j=1}^{\infty} \cov({\chi}_{jk},{\chi}_{jk'})=0$ and $\sum_{j=1}^{\infty} T_{n}(j,k,j,k')=0$.

\newpage
\setcounter{page}{1}
\bibliographystyle{apa-good}
\bibliography{kehui081016}
\end{document}